\documentclass[twocolumn]{aastex6}
\usepackage{apjfonts} 
\usepackage{amsmath} 

\usepackage{natbib}

\newcommand{\re}{$R_{\oplus}$}
 
\newcommand{\teff}{$T_{\rm eff}$}

\newcommand{\ct}{\citet}

\begin{document}
\title{Observing the Atmospheres of Known Temperate Earth-sized Planets with JWST}

\author{Caroline V. Morley\altaffilmark{1,2}, Laura Kreidberg\altaffilmark{1,3}, Zafar Rustamkulov\altaffilmark{4}, Tyler Robinson\altaffilmark{4,5}, Jonathan J. Fortney\altaffilmark{4}}
\date{August 2017}
\shorttitle{TRAPPIST and MEarth with JWST}
\shortauthors{Morley et al.}

\altaffiltext{1}{Department of Astronomy, Harvard University, caroline.morley@cfa.harvard.edu} 
\altaffiltext{2}{Sagan Fellow} 
\altaffiltext{3}{Harvard Society of Fellows} 
\altaffiltext{4}{Department of Astronomy \& Astrophysics, University of California Santa Cruz} 
\altaffiltext{5}{Department of Physics \& Astronomy, Northern Arizona University} 

\begin{abstract}

Nine transiting Earth-sized planets have recently been discovered around nearby late M dwarfs, including the TRAPPIST-1 planets and two planets discovered by the MEarth survey, GJ 1132b and LHS 1140b. These planets are the smallest known planets that may have atmospheres amenable to detection with JWST. We present model thermal emission and transmission spectra for each planet, varying composition and surface pressure of the atmosphere. We base elemental compositions on those of Earth, Titan, and Venus and calculate the molecular compositions assuming chemical equilibrium, which can strongly depend on temperature. Both thermal emission and transmission spectra are sensitive to the atmospheric composition; thermal emission spectra are sensitive to surface pressure and temperature. We predict the observability of each planet's atmosphere with \emph{JWST}. GJ 1132b and TRAPPIST-1b are excellent targets for emission spectroscopy with JWST/MIRI, requiring fewer than 10 eclipse observations. Emission photometry for TRAPPIST-1c requires 5--15 eclipses; LHS 1140b and TRAPPIST-1d, TRAPPIST-1e, and TRAPPIST-1f, which could possibly have surface liquid water, may be accessible with photometry. Seven of the nine planets are strong candidates for transmission spectroscopy measurements with \emph{JWST}, though the number of transits required depends strongly on the planets' actual masses. Using the measured masses, fewer than 20 transits are required for a 5$\sigma$ detection of spectral features for GJ 1132b and six of the TRAPPIST-1 planets. Dedicated campaigns to measure the atmospheres of these nine planets will allow us, for the first time, to probe formation and evolution processes of terrestrial planetary atmospheres beyond our solar system.

\end{abstract}


\maketitle

\section{Introduction}\label{introduction}

Planets smaller than Neptune are now known to be common around both Sun-like stars and cooler M dwarfs \citep{Petigura13, Dressing15, Burke15}, with the \emph{Kepler} mission revealing thousands of planet candidates around distant stars \citep{Borucki11, Coughlin16}. As we move toward understanding those planetary systems, one of the key next steps is to measure the compositions of planets spanning from Jupiter-size to Earth-size and smaller, searching for trends that allow us to determine the formation and evolution histories of planets in the Milky Way. In particular, measuring the abundances of molecular species in the atmospheres of a diverse range of planets may reveal how the atmosphere was accreted or outgassed, and subsequently evolved \citep{Oberg11, Espinoza17, Booth17}. 

One of the most challenging classes of planets to characterize are the smallest planets, around the size of the Earth and smaller. These planets are a challenge largely because of their small transit depths, decreasing the precision achievable for transit and eclipse observations. While Neptune and giant planet atmospheres will be detectable around Sunlike stars with the upcoming \emph{James Webb Space Telescope} (\emph{JWST}) \citep[e.g.,][]{Greene16}, the atmospheres of planets around the size of Earth will not be. 

However, the atmospheres of small planets orbiting some of the smallest, closest stars will be readily detectable. To that end, a number of surveys have been searching for planets around these nearby stars \citep{Nutzman08b, Jehin11, Ricker14, Howell14}, and to date nine Earth-sized planets have been detected that are amenable to atmospheric characterization. These planets are likely to be extensively studied with \emph{JWST} and what we learn will form the basis of our knowledge about Earths outside the solar system for the next decades, until the launch of large aperture telescopes capable of directly imaging Earths around Sunlike stars. 

\begin{figure}[t]
\center \includegraphics[width=3.7in]{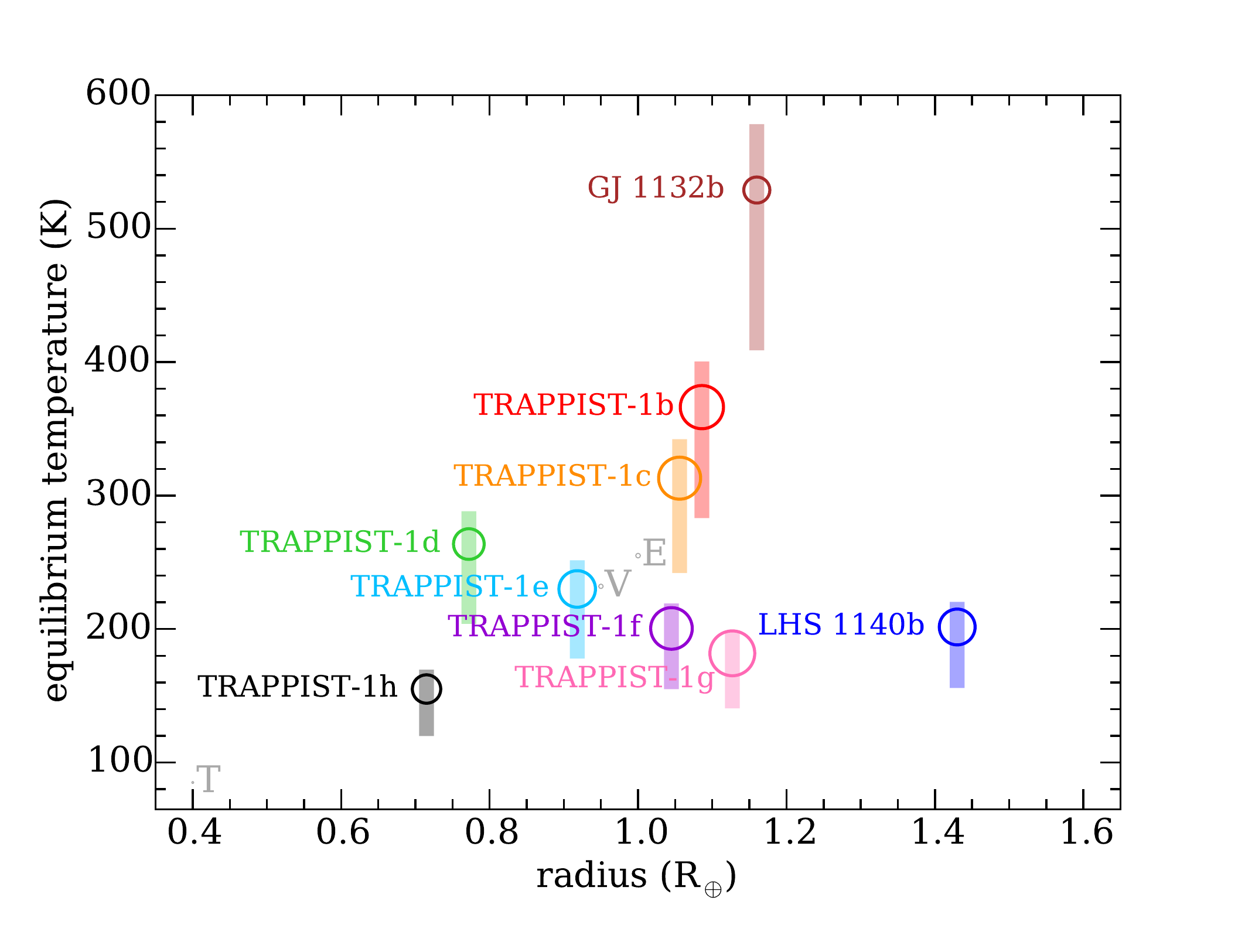}
 \caption{The radii and equilibrium temperature ranges of the 9 planets considered in this study. The radius of the marker is scaled proportionally with the radius of the planet divided by the radius of the star. The marker is centered on the equilibrium temperature assuming a Bond albedo of 0.3; the extent of the transparent colored bar for each planet shows the equilibrium temperature range from the lowest T$_{eq}$ (assuming a Venus-like Bond albedo of 0.7) to the highest T$_{eq}$ (assuming a Bond albedo of 0.0). The locations of Earth, Venus, and Titan are shown in light gray (labeled E, V, and T), assuming Bond albedos of 0.31, 0.75, and 0.2 respectively. The marker size of these is scaled with the ratio of the planet radius to the Sun's radius; the markers are therefore hard to see since the Sun is much larger than the M dwarf exoplanet hosts considered here.}
\label{planet_summary}
\end{figure}

\subsection{9 Earth-sized Planets Orbiting Small Stars}

During the past two years, nine planets close to Earth in size have been discovered around nearby M dwarfs cooler than 3300 K. These planets include the 7 planets in the TRAPPIST-1 system and two planets discovered by the MEarth survey, GJ 1132b and LHS 1140b \citep{Gillon16,Gillon17, BertaThompson15, Dittmann17}. These planets are the smallest planets discovered to date that will be amenable to atmospheric characterization with \emph{JWST}. 

The properties of these planets are summarized in Figure \ref{planet_summary}. They span equilibrium temperatures from $\sim$130 K to $\gtrsim$500 K, and radii from 0.7 to 1.43 \re. Some of these planets orbit at distances amenable to surface liquid water, though the surface temperature will depend strongly on the albedo of the planet and the thickness and composition of its atmosphere.

The host stars of these three systems differ considerably from each other, creating a test bed for measuring how the environment shapes planetary atmospheres. For example, TRAPPIST-1 is a very cool (2550 K) star just barely massive enough to fuse hydrogen into helium and be classified as a star. It is also a fairly active star: \citet{Wheatley17} used XMM-Newton to measure the X-ray emission of TRAPPIST-1 and found that it is a relatively strong X-ray source, with X-ray luminosity about that of the quiet Sun, despite a $\sim$2000 times lower bolometric luminosity. This high energy emission will affect the chemistry of the atmosphere and potentially drive mass loss, which would vary in strength with orbital distance. LHS 1140b is slightly hotter than TRAPPIST-1, with a temperature of 3100 K. In contrast to the TRAPPIST-1 star, LHS 1140 and GJ 1132 are both slow rotators (131 day and 125 day rotation periods, respectively), indicating that they are likely old, and therefore relatively inactive \citep{Dittmann17, BertaThompson15}. Since there are planets around each star with similar bolometric incident flux, these systems provide a testbed for determining the effect of high energy irradiation on planetary atmospheres.

\subsection{The Atmospheres of Earth-sized Planets}

The atmospheres of the terrestrial planets in our own solar system have a diverse set of properties, from thick cloudy atmospheres made mostly of carbon dioxide like Venus', to Earth's biotically-altered nitrogen-oxygen atmosphere, to Titan's thick nitrogen-methane atmosphere complete with a methane hydrological cycle, to seasonally-varying thin atmospheres of Mars and Pluto \citep[e.g.,][]{Martinez17}. These atmospheres are each shaped by the planet's formation and evolution, including the accretion of volatiles early in the planet's lifetime, differentiation of the planet's interior, outgassing of a secondary atmosphere, and subsequent mass loss from the atmosphere as the planet evolves \citep[e.g.,][]{Pierrehumbert10, ChambHunt}. The atmospheres of the terrestrial exoplanets we study around cool stars will undoubtedly be even more diverse than those in the solar system \citep{Meadows10, Wordsworth14}. 

Previous work has examined the environment, assessed potential habitability, and modeled the spectra of some of the nine planets we study here. In particular, \citet{Barstow16} generated synthetic spectra for the inner 3 TRAPPIST-1 planets and determined that, if present, Earth-like levels of ozone would be detectable with 60 transits for TRAPPIST-1b and 30 transits for TRAPPIST-1c and TRAPPIST-1d. \citet{Schaefer16} studied the possible compositions of GJ 1132b given a magma ocean and water vapor atmosphere as the initial condition, finding that such atmospheres are susceptible to considerable mass loss, which can cause O$_2$ to build in abundance. Though Proxima b does not transit so we do not include it in this work, other works have examined possible atmospheres of Proxima b \citep[e.g.,][]{Meadows16} and their detectability \citep[e.g.,][]{Kreidberg16, Luger17}.  

\citet{Wolf17} assessed the relative habitability of each of the TRAPPIST-1 planets using a climate model (National Center for Atmospheric Research Community Atmosphere Model), and find that TRAPPIST-1e is the most amenable to liquid water on the surface. In contrast, using a climate-vegetation energy balance model, \citet{Alberti17} find that TRAPPIST-1d is most amenable to surface liquid water. \citet{Turbet17} instead used a 3D global climate model to explore climate, atmospheric condensation and photochemistry for the TRAPPIST-1 planets; they find that TRAPPIST-1e is the best candidate for habitability.

In this work, we build a catalog of thermal emission and transmission spectra for a broad variety of potential atmospheric compositions. The compositions chosen here remain agnostic to the complex physics and chemistry that would be necessary to predict \emph{a priori} the atmospheric compositions from first principles given chosen initial conditions. Instead, we choose three very different atmospheric compositions that are by definition possible outcomes of planet formation simply because they exist in our own solar system: those of Earth, Venus, and Titan. We use the \emph{elements} that compose these atmospheres, and we calculate their molecular compositions given the different incident flux levels for each of the 9 planets. We model the thermal emission and transmission spectra of these planets for a broad range of surface pressures and albedos. We then use these model atmospheres to predict the observability of these planets with \emph{JWST}.

\subsection{Structure of this Work}

In Section \ref{methods}, we describe how we construct these model atmospheres and calculate synthetic spectra. In Section \ref{modelresults}, we show the chemistry, opacities, surface temperatures, and model spectra; in Section \ref{jwstresults} we show results from \emph{JWST} simulations and assess the observability of each planet. In Section \ref{discuss} we discuss the results including additional parameters that should be assessed in future work, and finally in Section \ref{conclude} we summarize our conclusions.

\begin{figure}[t]
\center \includegraphics[width=3.7in]{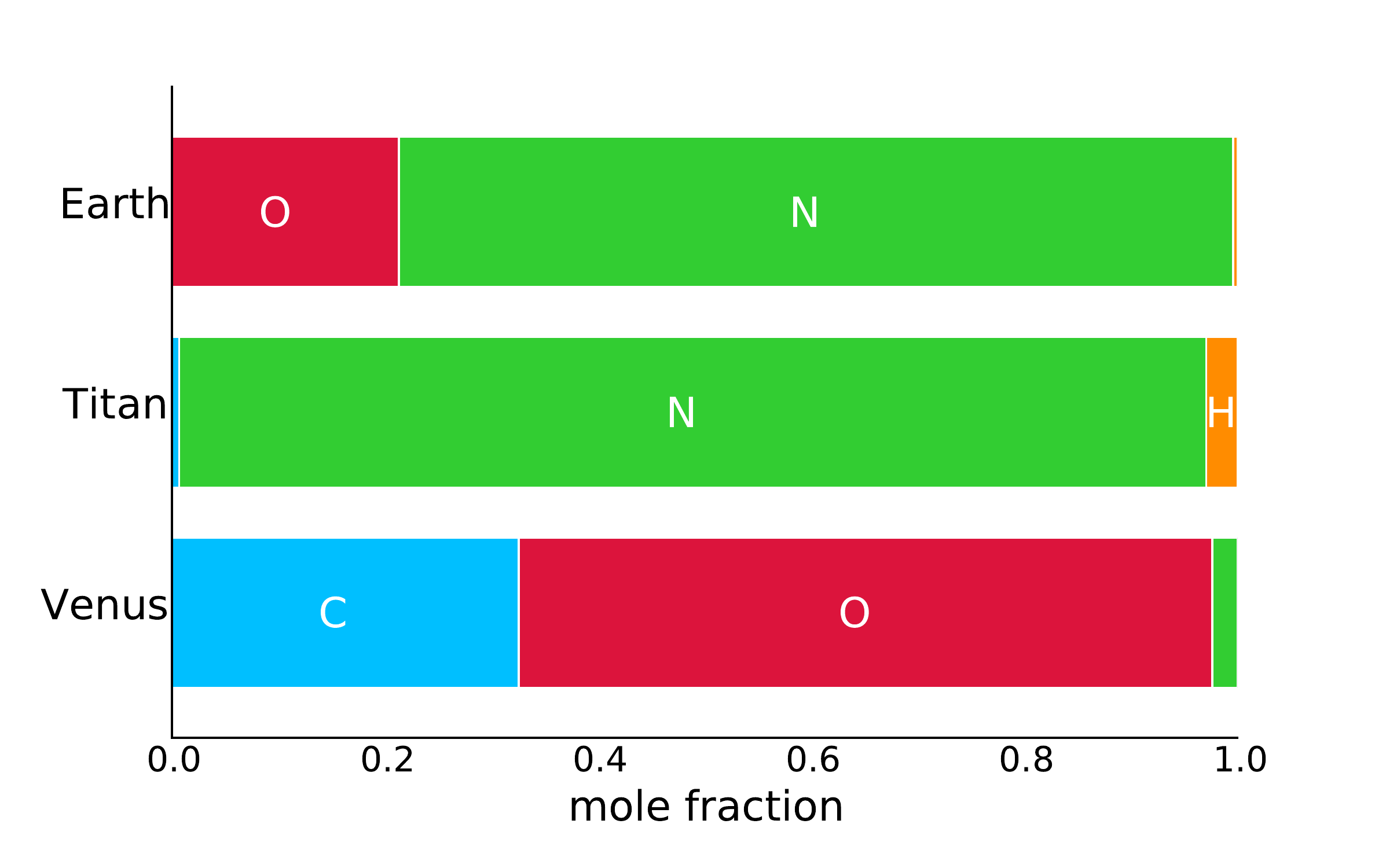}
\vspace{-0.5in}
\center \includegraphics[width=3.7in]{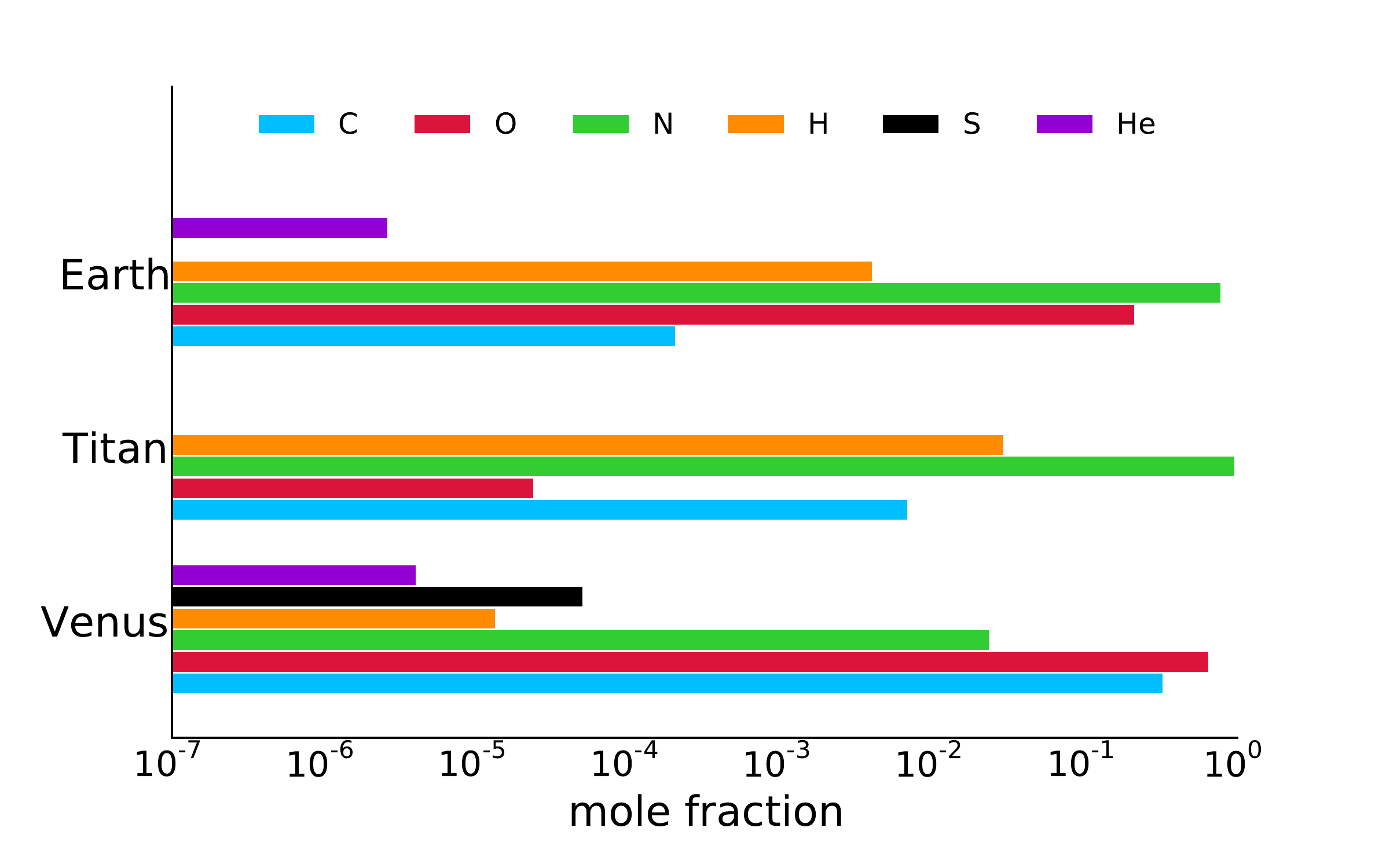}
 \caption{Compositions of the solar system planets Earth, Titan, and Venus. The top panel shows a stacked bar graph with each of the major elements that compose each atmosphere. The bottom panel shows the same information using a log scale to show the minor constituents. }
\label{elements}
\end{figure}

\section{Methods} \label{methods}

\subsection{Planet Properties}

We take the majority of properties used to model these nine planets from the discovery papers \citep{Gillon17, BertaThompson15, Dittmann17}, including planetary radii, stellar radii and temperatures, and orbital distance.  Many of the masses of these small planets are poorly constrained; because this has a strong impact on transmission spectra, we calculate transmission spectra using two masses for each planet: the central value of the measured mass and a mass derived from an empirical mass-radius relationship. We use an empirical mass--radius relationship from \citet{Weiss14}, which is consistent with a rocky terrestrial composition \citep{Zeng16}. We use measured masses from the discovery papers for GJ 1132b, LHS 1140b, and TRAPPIST-1f and from \citet{Wang17} for the remaining TRAPPIST-1 planets. In thermal emission we consider only one mass for each planet because thermal emission is less sensitive to surface gravity, using the measured masses for GJ 1132b, LHS 1140b, and TRAPPIST-1f and the mass derived from the \citet{Weiss14} relationship for the other planets. 

\begin{figure}[t]
\center \includegraphics[width=3.7in]{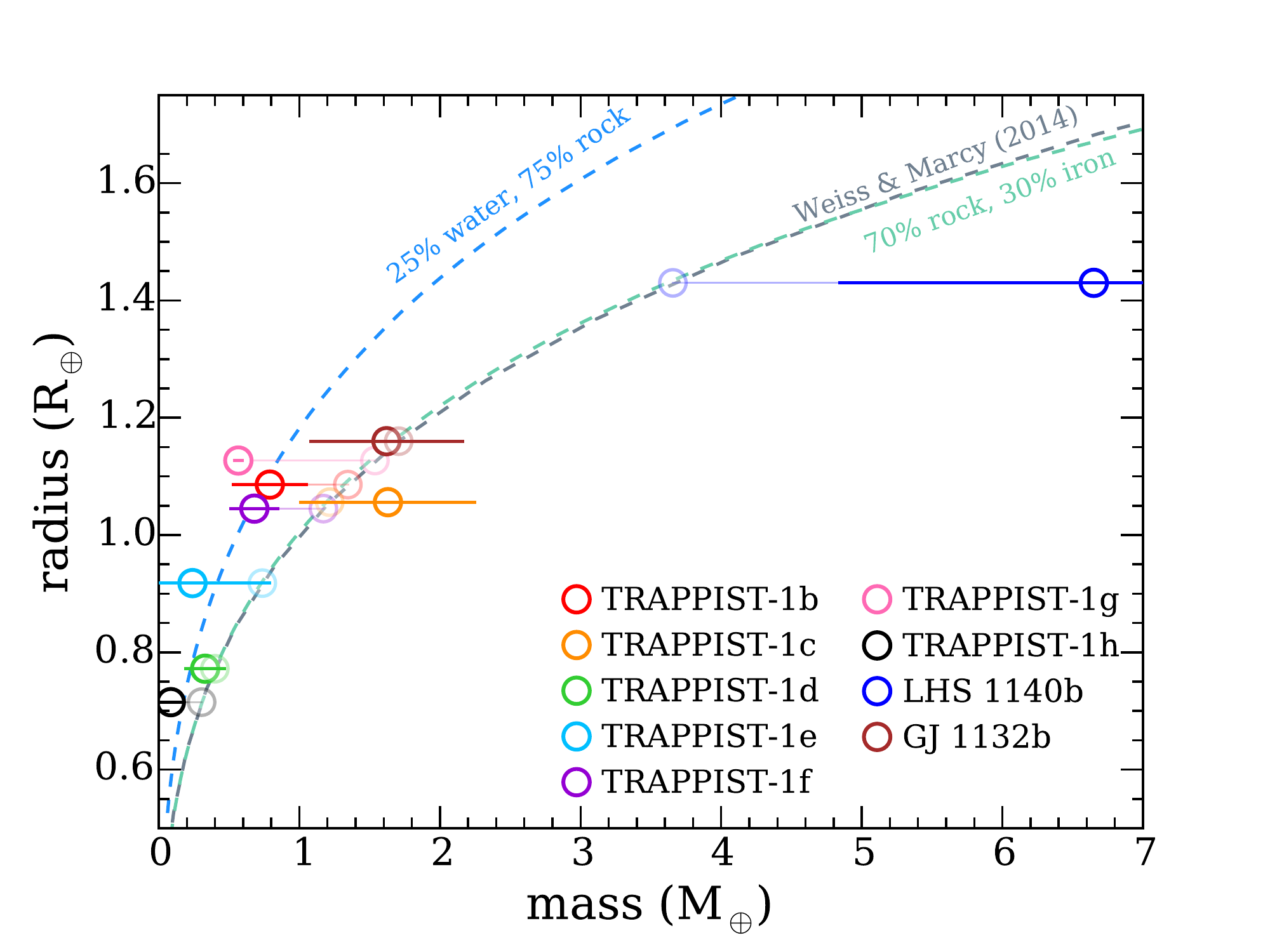}
 \caption{The masses and radii of the planets considered here, with empirical and theoretical mass-radius relationships. Each planet is shown as an open circle for both the measured mass (solid circle), and for an empirical Earth-like compsitions (transparent circle). The two masses are connected by a transparent line; the error bar on the measured mass is shown as a solid line. The mass-radius relationships are shown as dashed lines and include an iron/rock and water/rock theoretical relationship \citep{Zeng16} and an empirical mass-radius relationship \citep{Weiss14}.  }
\label{planet_masses}
\end{figure}

\subsection{Elemental Compositions of 3 Solar System Terrestrial Planets}

We consider here the compositions of three of the solar system's atmospheres, which represent three different outcomes of the planet formation and evolution process around a Sunlike star. We do not imagine that these are the only possible compositions, but use these as a starting point to model how disparate planetary histories might manifest in the compositions of these exoplanets, and to study the observability of these different scenarios. 

In the solar system, the atmospheric compositions of the terrestrial planets are controlled by the accretion of volatiles at early ages, followed by the outgassing and escape of volatiles over the lifetime of the planet. These processes are a function of both the chemistry of the planetary interior and the physics of the interaction between sunlight and the atmosphere. For example, Earth's interior segregated early in its lifetime, leaving an iron-poor mantle, which allowed oxygen to interact with other elements. The oceans play a major role in controlling the composition of Earth's atmosphere through carbon-silicate weathering; in this process, Earth sequesters most of its CO$_2$ underground in limestone rocks, leaving an atmosphere rich in N$_2$. On Titan, its ice-rich mantle may release NH$_3$ and CH$_4$, allowing Titan to maintain its atmosphere \citep{Horst17}.  Meanwhile Venus, at slightly higher irradiation from the Sun than the Earth, is believed to have experienced a `runaway greenhouse' \citep{Ingersoll69}, where the planet cannot effectively radiate its absorbed solar energy, which leads to rapid heating to high surface temperature \citep{Goldblatt15}. Photo-dissociation of water followed by the preferential loss of hydrogen to space caused Venus to become dry, leaving a mostly CO$_2$ atmosphere. 

The compositions of the atmospheres of Earth \citep{Wallace77}, Titan \citep{Niemann05, Horst08}, and Venus \citep{Basilevsky03, Bertaux07} are shown in Figure \ref{elements}. The panels show the same information, on a linear scale in the top panel (to illustrate the differences in bulk properties) and log scale in the bottom panel (to show species with lower abundances). Earth's atmosphere is predominantly nitrogen and oxygen with some carbon, hydrogen, and helium. Venus's atmosphere is mostly C and O (as CO$_2$) with sulfur, nitrogen, and helium. Titan, like Earth, has a mostly nitrogen atmosphere, with hydrogen, carbon, and oxygen. We ignore the noble gases, which are present in all terrestrial atmospheres but as inert gases are a challenge to detect remotely and have no effect on the atmospheric chemistry. 

We ignore the effect of changing the surface temperature on the atmospheric chemistry. For example, if the surface temperature on Earth increased enough, the oceans would evaporate, increasing the hydrogen and oxygen abundances. Likewise, the higher surface temperature of Venus causes more sulfur to vaporize, so a cooler planet would likely have a lower sulfur abundance. We do not model these processes self-consistently here, instead keeping the ratios of elements consistent between models.

\subsection{Constructing Simple Model Atmospheres} \label{makin_atm}

\begin{figure}[t]
\center \includegraphics[width=3.7in]{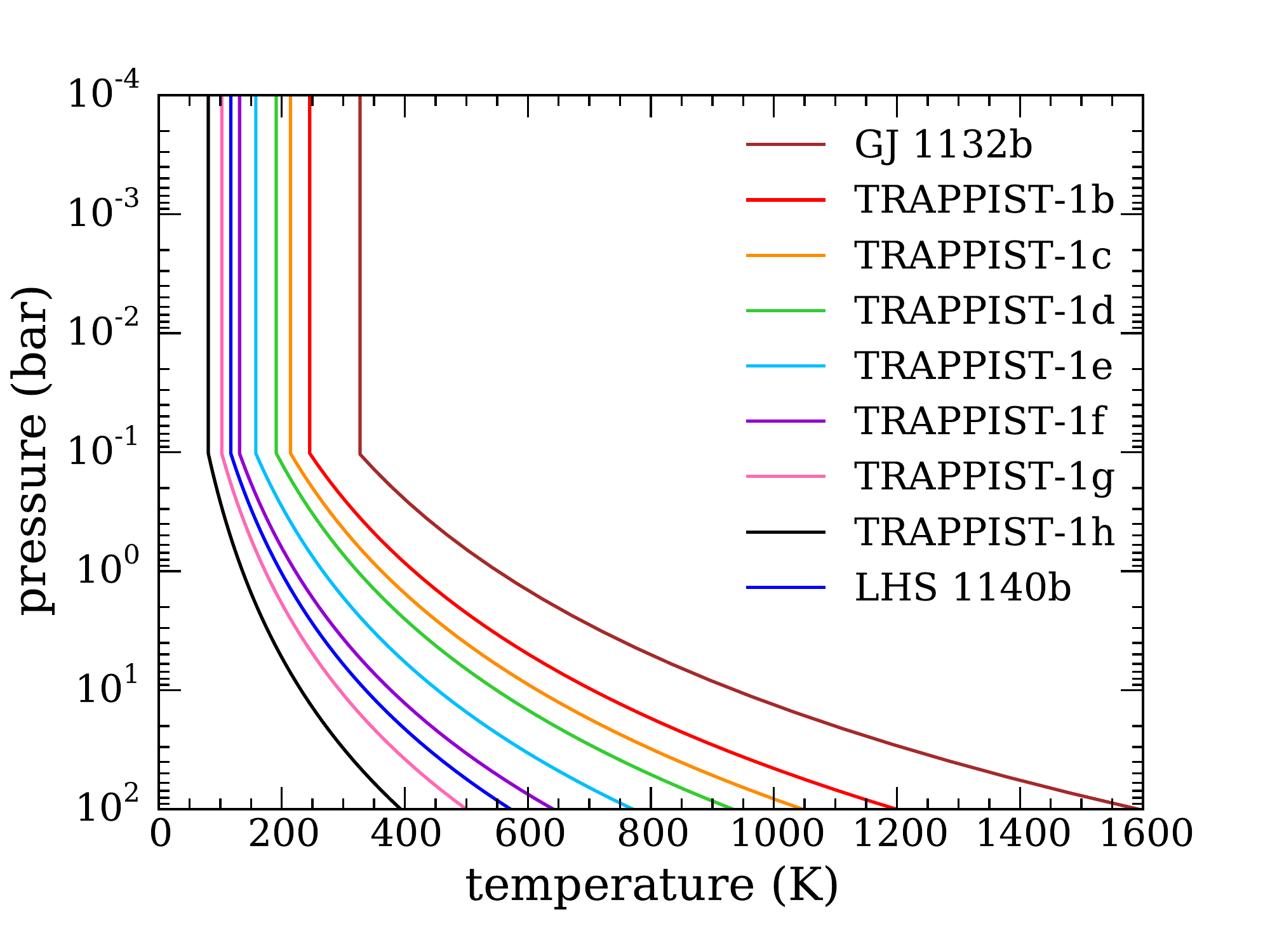}
 \caption{Examples of model pressure-temperature profiles for each planet. All models shown have a surface pressure of 100 bar and Venus-based composition. The lower part of the atmosphere follows a dry adiabat (assumed to be 100\% CO$_2$), while above the 0.1 bar `tropopause' the model atmospheres are isothermal.  }
\label{planet_pt}
\end{figure}

We calculate temperature structures for each planet using analytic formulae. This simplification is adequate for the broad brush analysis we are doing here, but more detailed modeling could implement, e.g., radiative-convective equilibrium models to calculate the temperature structure. 

We model a range of surface pressures from very tenuous atmospheres (10$^{-4}$ bar, which is between Mars' and Pluto's) to thick Venus-like atmospheres (100 bar) in increments of 1 dex. For atmospheres above 1 bar in surface pressure, we use dry adiabatic profiles from the surface to 0.1 bar, where the solar system planets' tropopauses typically form \citep{Robinson14c}; at <0.1 bar pressures, we assume an isothermal pressure-temperature (P--T) profile. For atmospheres with surface pressures 0.1 bar or less, we assume a dry adiabatic profile from the surface to the layer in which the temperature is equal to the skin temperature, defined as $T_{skin}=\frac{1}{2^{1/4}}T_{rad}$, where $T_{rad}$ is the equivalent temperature to the energy that the planet radiates to space \citep{Pierrehumbert10}. Above this `skin layer' we again assume an isothermal P--T profile. For all adiabatic profiles we assume an N$_2$ atmosphere (for Titan and Earth-based compositions) and a CO$_2$ atmosphere (for Venus-based compositions), and we neglect the effect of condensation on P-T profiles. Examples of these model P--T structures are shown in Figure \ref{planet_pt}. We assume that the planetary surface has a uniform albedo (and therefore uniform emissivity) equal to that of the model planet.

We choose the surface temperature ($T_{\rm surf}$) for each surface pressure ($P_{\rm surf}$) iteratively by calculating spectra for a range of different $T_{\rm surf}$ for each planet/pressure/composition combination. We then iterate $T_{\rm surf}$ to balance incoming radiation with outgoing radiation. That is, we find a $T_{\rm surf}$ for which the effective temperature ($T_{\rm eff}$) of the model planet is equal to the equilibrium temperature ($T_{\rm eq}$) for that planet (given a particular Bond albedo). Bond albedo for a particular planet will be a strong function of both its geometric albedo and of the incident stellar spectrum. For this study, we focus on the limiting Bond albedos of 0.0 (fully absorbing), 0.3 (Earth-like albedo), and 0.7 (Venus-like albedo, and the largest Bond albedo of a body in the solar system).  

We calculate the gas abundances assuming chemical equilibrium along each profile. We use the Chemical Equilibrium with Applications model (CEA, Gordon \& McBride 1994) to compute the thermochemical equilibrium molecular mixing ratios (with applications to exoplanets see, \ct{Visscher10, Line10, Moses11, Line11, Line13b}). CEA minimizes the Gibbs free energy with an elemental mass balance constraint given a local temperature, pressure, and elemental abundances. We include molecules containing H, C, O, N, S, and He.

\begin{figure}[t]
\center \includegraphics[width=3.4in]{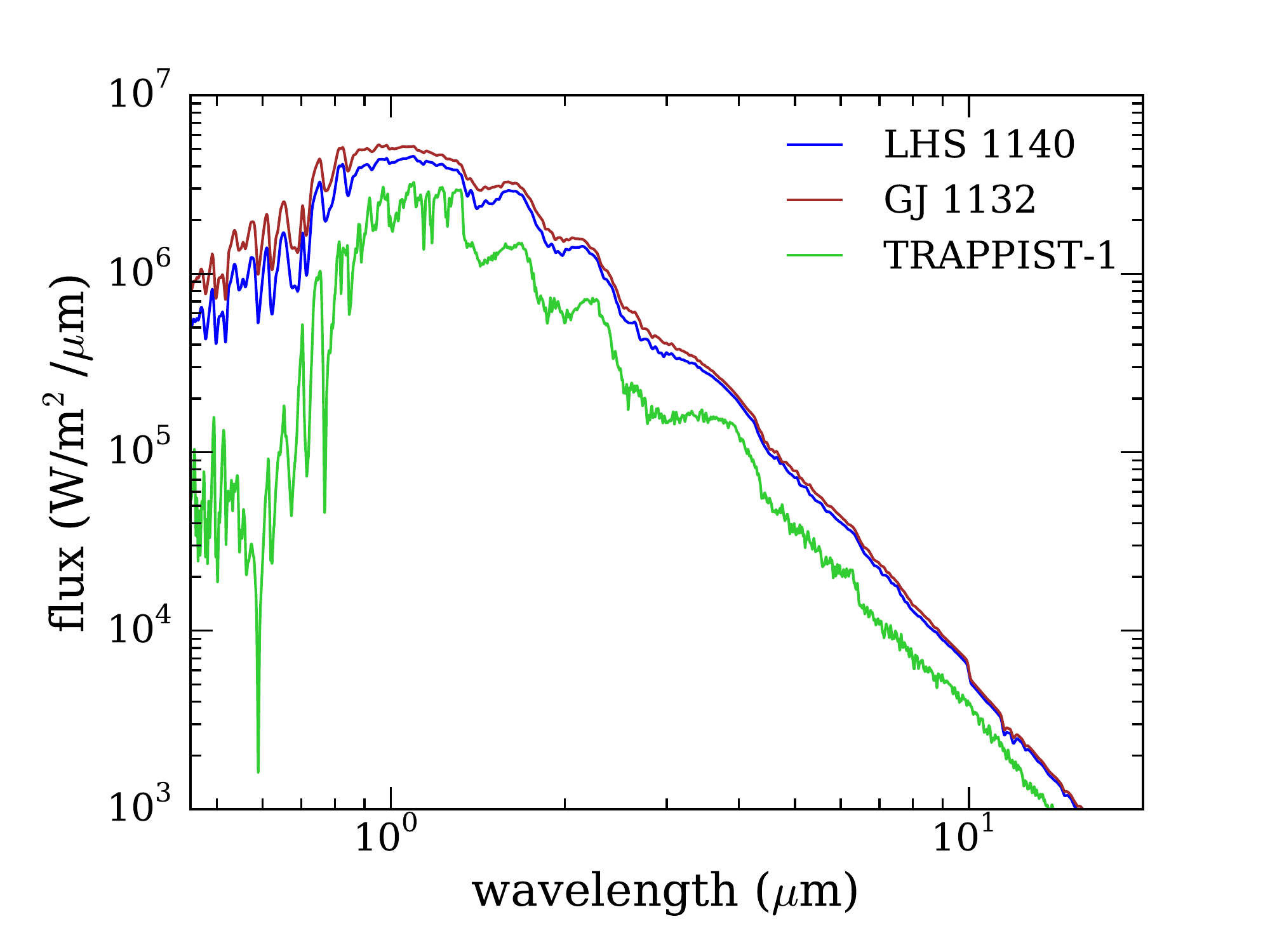}
 \caption{Spectra of the 3 M dwarf host stars. LHS 1140 is an M4.5 dwarf (\teff$\sim$3130 K); GJ 1132 is an M3.5 dwarf (\teff$\sim$3270 K); TRAPPIST-1 is an M8 dwarf (\teff$\sim$2550 K). The flux is calculated at the surface of the star. Note the significant deviations from a blackbody-like spectrum, with water absorption features as well as metal oxides and hydrides. }
\label{stars}
\end{figure}

\subsection{Thermal Emission Spectra} \label{thermal_model}

We calculate thermal emission spectra using the model described in the Appendix of \citet{Morley15}. We calculate the thermal emission emerging from the top of an atmosphere of arbitrary composition using the C version of the open-source radiative transfer code \texttt{disort} \citep{Stamnes88, Buras11}. \texttt{disort} is a numerical implementation of the discrete-ordinate method for monochromatic (unpolarized) radiative transfer, including absorption, emission, and scattering, in non-isothermal, vertically inhomogeneous media. 

In order to calculate the emergent spectrum, we calculate the optical depth $\tau$, single-scattering albedo $\omega$, asymmetry factor $g$, and temperature $T$ using the results from the calculations described in Section \ref{makin_atm}. The only scattering process we consider in this work is Rayleigh scattering. 

The opacity database is based on \citet{Freedman08} with updates described in \citet{Freedman14}, including methane \citep{Yurchenko14} and carbon dioxide \citep{Huang13, Huang14}. For these emission spectra, we include line lists for the following molecules: CO$_2$, H$_2$O, CH$_4$, CO, NH$_3$, H$_2$S, C$_2$H$_2$, C$_2$H$_4$, C$_2$H$_6$, HCN, O$_2$, O$_3$, NO, NO$_2$, and SO$_2$. We include collision-induced opacities (CIA) from N$_2$-N$_2$, O$_2$-O$_2$, CO$_2$-CO$_2$, and N$_2$-CH$_4$ \citep{Gruszka97, Baranov04, Lee16, Bezard90, Wordsworth10, Borysow93, Lafferty96, Hartmann17}, as well as hydrogen and helium CIA opacities \citep{Richard12}.

\subsection{Transmission Spectra}

We calculate transmission spectra using a new model transmission spectrum code described in more detail in the Appendix. In short, this model uses the optical depth calculations presented in \citet{Morley15} and the matrix prescription for calculating transmission spectra presented in \citet{Robinson17}. Opacities are as described in Section \ref{thermal_model}.

\begin{figure}[thb]
\center \includegraphics[width=3.4in]{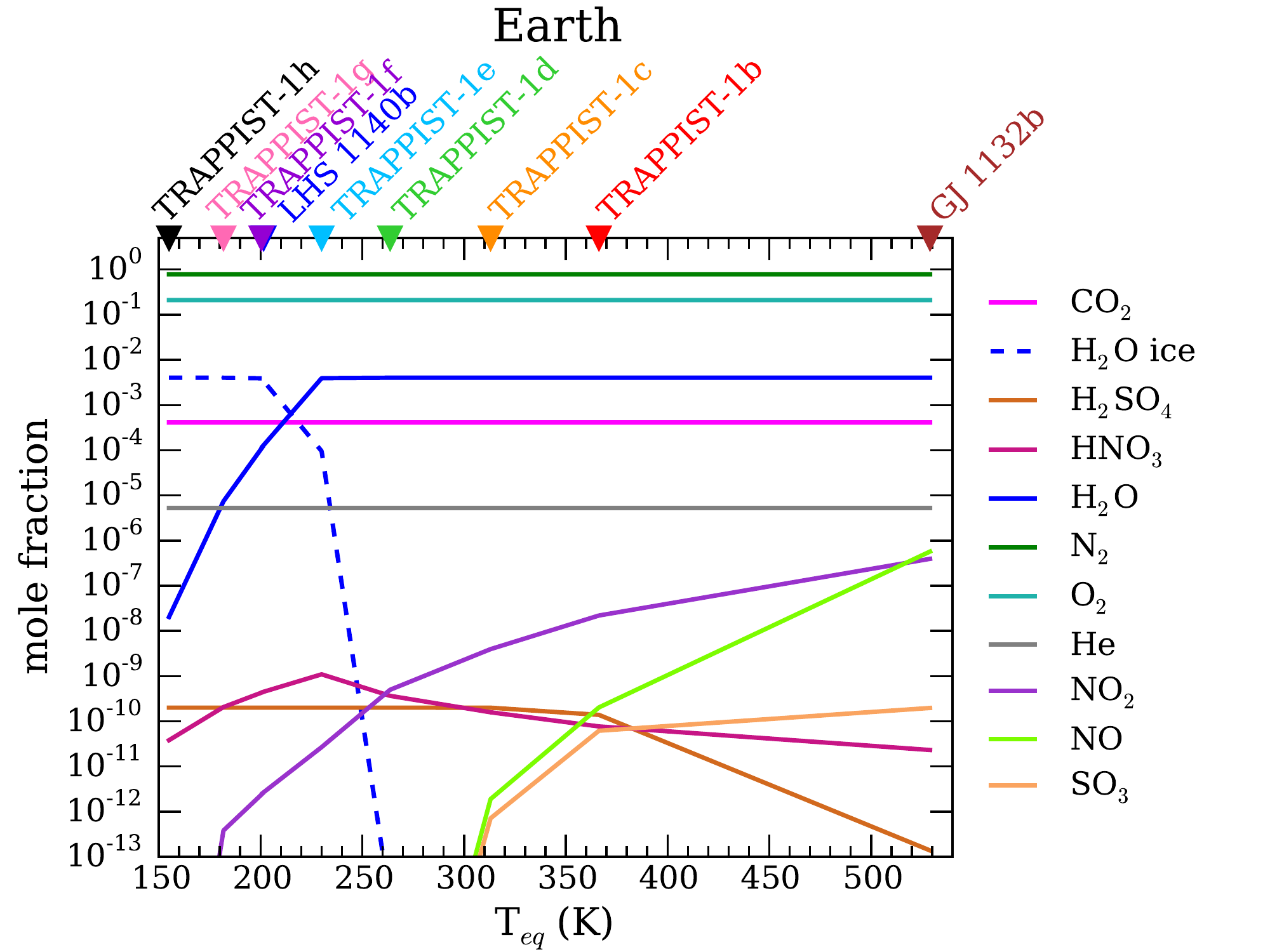}
\vspace{-0.3in}
\center \includegraphics[width=3.4in]{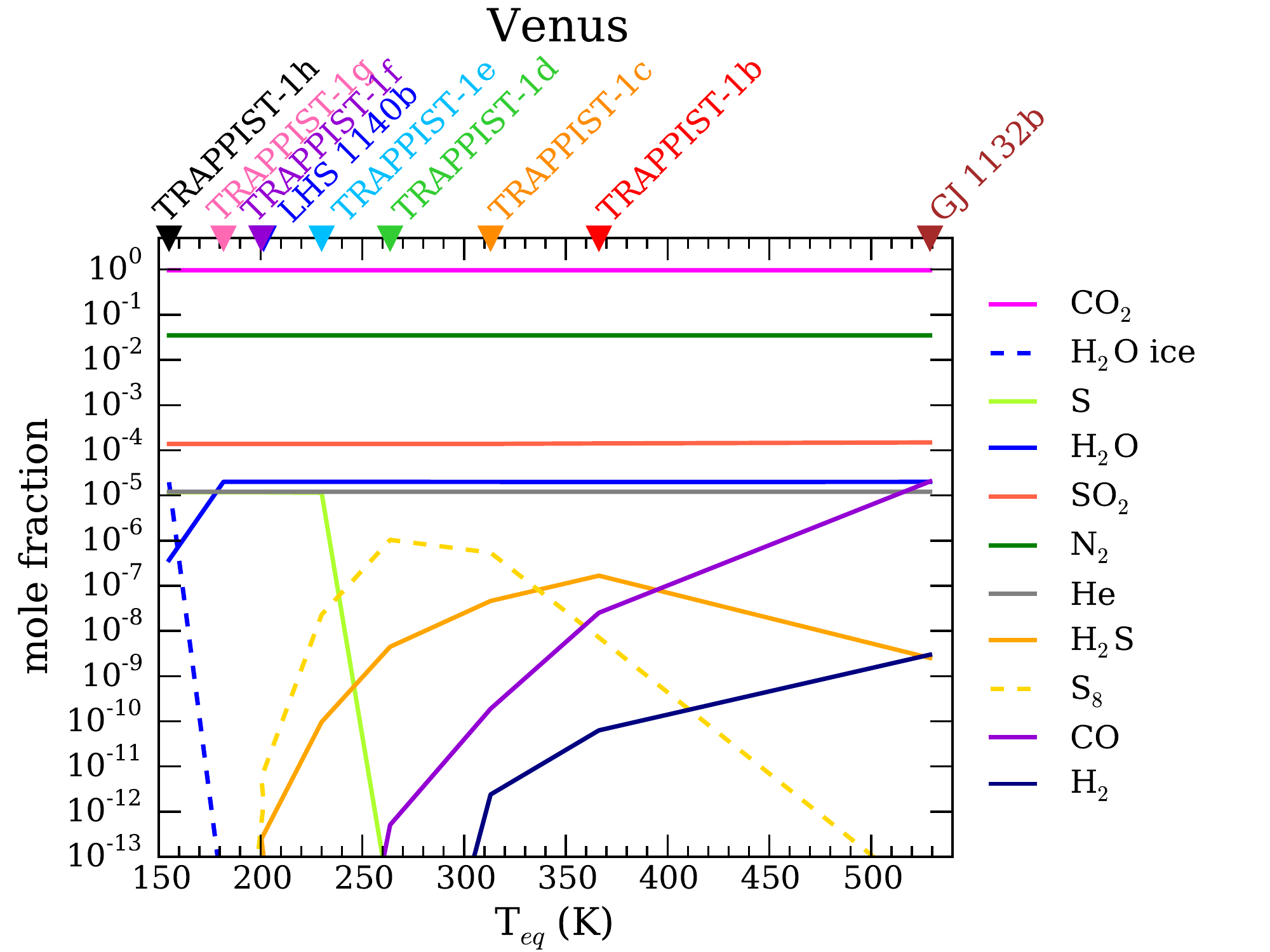}
\vspace{-0.3in}
\center \includegraphics[width=3.4in]{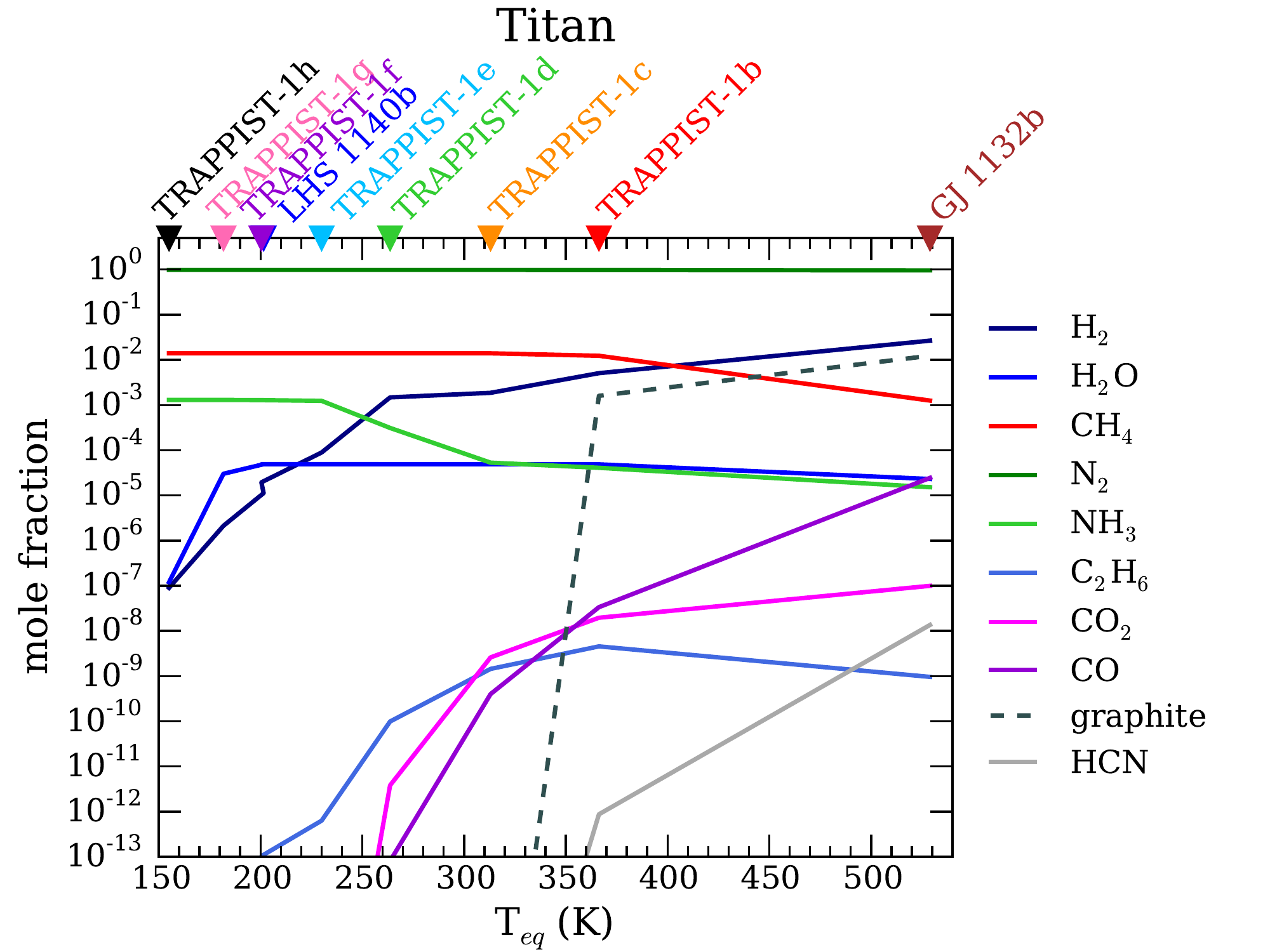}
 \caption{The abundances of different species versus equilibrium temperature of each model is shown. For the equilibrium temperature of each planet (shown at top), the mole fraction of the species at the surface is shown. Each panel shows a different elemental composition (Earth, Titan, and Venus from top to bottom); all have surface pressures of 1 bar.  }
\label{chemistry_synopsis}
\end{figure}

\subsection{Stellar Spectra}

Spectra for the three host stars are shown in Figure \ref{stars}. LHS 1140 and GJ 1132 spectra are calculated using PHOENIX NextGen models \citep{Allard97, Baraffe97, Baraffe98, Hauschildt99}. The TRAPPIST-1 spectrum is calculated for this paper using the radiative-convective atmosphere models described in \citet{McKay89, Marley96, Saumon08, Fortney08b, Morley15}, assuming a dust-free model atmosphere.

\section{Modeling Results} \label{modelresults}

In this section, we will present chemistry, opacities, and spectra for model atmospheres with the compositions, surface pressures, and Bond albedos described in Section \ref{methods}. The model spectra are all publically available online both at the first author's website\footnote{www.carolinemorley.com/models} and permanently linked at a data repository\footnote{https://doi.org/10.5281/zenodo.1001033}. 

\subsection{Chemistry}

Figure \ref{chemistry_synopsis} summarizes the abundances of various molecular species in chemical equilibrium for models with the equilibrium temperatures of each planet, ranging from 150 K (TRAPPIST-1h) to over 500 K (GJ 1132b). The plotted examples are shown for atmospheres with surface pressures of 1 bar, and the abundances shown are at the surface. 

For Earth-based elemental composition (top panel), the atmosphere consists of nitrogen (N$_2$) and oxygen (O$_2$) regardless of atmospheric temperature. Water vapor (H$_2$O) or water ice is the next-most abundant species for the warmer models (T$_{eq}$>200 K) and colder models (T$_{eq}$<200 K) respectively, followed by carbon dioxide (CO$_2$) and helium (He). The trace species in the atmosphere change with temperature; for example, nitric oxide (NO) and nitrogen dioxide (NO$_2$) increase in abundance with temperature, while sulfuric acid (HSO$_4$) and nitric acid (HNO$_3$) decrease. 

For Titan-based elemental composition (middle panel), the atmosphere consists mainly of nitrogen (N$_2$). The other species vary in abundance as a function of temperature. For cooler objects (T$_{eq}$<250 K), carbon is in the form of methane (CH$_4$), while for warmer objects carbon condenses into (solid) graphite and H$_2$ becomes the second-most abundant gas. Ammonia (NH$_3$) is present in relatively high abundances, as is H$_2$O. Carbon monoxide (CO) and hydrogen cyanide (HCN) both increase with temperature, and CO$_2$ and ethane (C$_2$H$_6$) are also present. 

For Venus-based elemental composition (bottom panel), the atmosphere is predominantly CO$_2$ and N$_2$. Sulfur dioxide (SO$_2$), H$_2$O, and He are the next-most abundant 3 species, and do not vary in abundance with temperature. The other trace species vary with temperature; monatomic S is present in cold atmospheres (T$_{eq}$<200 K), S$_8$ is present (as a solid) at intermediate temperatures, and hydrogen sulfide H$_2$S is present at somewhat warmer temperatures. Above T$_{eq}\sim$350 K, CO and H$_2$ become more abundant.

In general, the bulk atmospheric compositions predicted here match those of solar system Earth, Venus, and Titan. For example, Earth's atmosphere is mostly nitrogen and oxygen, with on average 0.4\% water vapor, 400 ppm of CO$_2$, and 5 of ppm He, very closely matching the equilibrium chemistry calculations for TRAPPIST-1e. However, the chemistry of the trace species is somewhat different; for example actual Earth has significant amounts of ozone (O$_3$, from photochemistry) and nitrous oxide (N$_2$O) and methane (from various sources including agriculture). Similarly, actual Venus has a 96.5\% CO$_2$ atmosphere, with 3\% N$_2$, 0.015\% SO$_2$, 0.002\% H$_2$O, and 0.0017\% CO, which essentially match the equilibrium chemistry predictions (for a 100 bar surface pressure). We do not consider a planet as cold as actual Titan here; for the coldest model we consider (TRAPPIST-1h), the nitrogen and methane abundances match those of Titan itself (98.4\% and 1.4\%), but we do note that the equilibrium models have considerable NH$_3$, not present in Titan. We discuss these differences is more detail in Section \ref{disequ}. Titan has a number of trace species that are not present at high abundance in the TRAPPIST-1h chemical equilibrium model, including C$_2$H$_2$ (5.5 ppm), C$_2$H$_6$ (20 ppm), H$_2$ (960 ppm), and CO (40 ppm), but Titan has less H$_2$O (<0.008 ppm) than the models. We note that at these cold temperatures, chemical reactions are often quite slow, so it is not surprising to deviate considerably from chemical equilibrium. 

\subsection{Opacities}

\begin{figure*}[h]
 \includegraphics[width=3.8in]{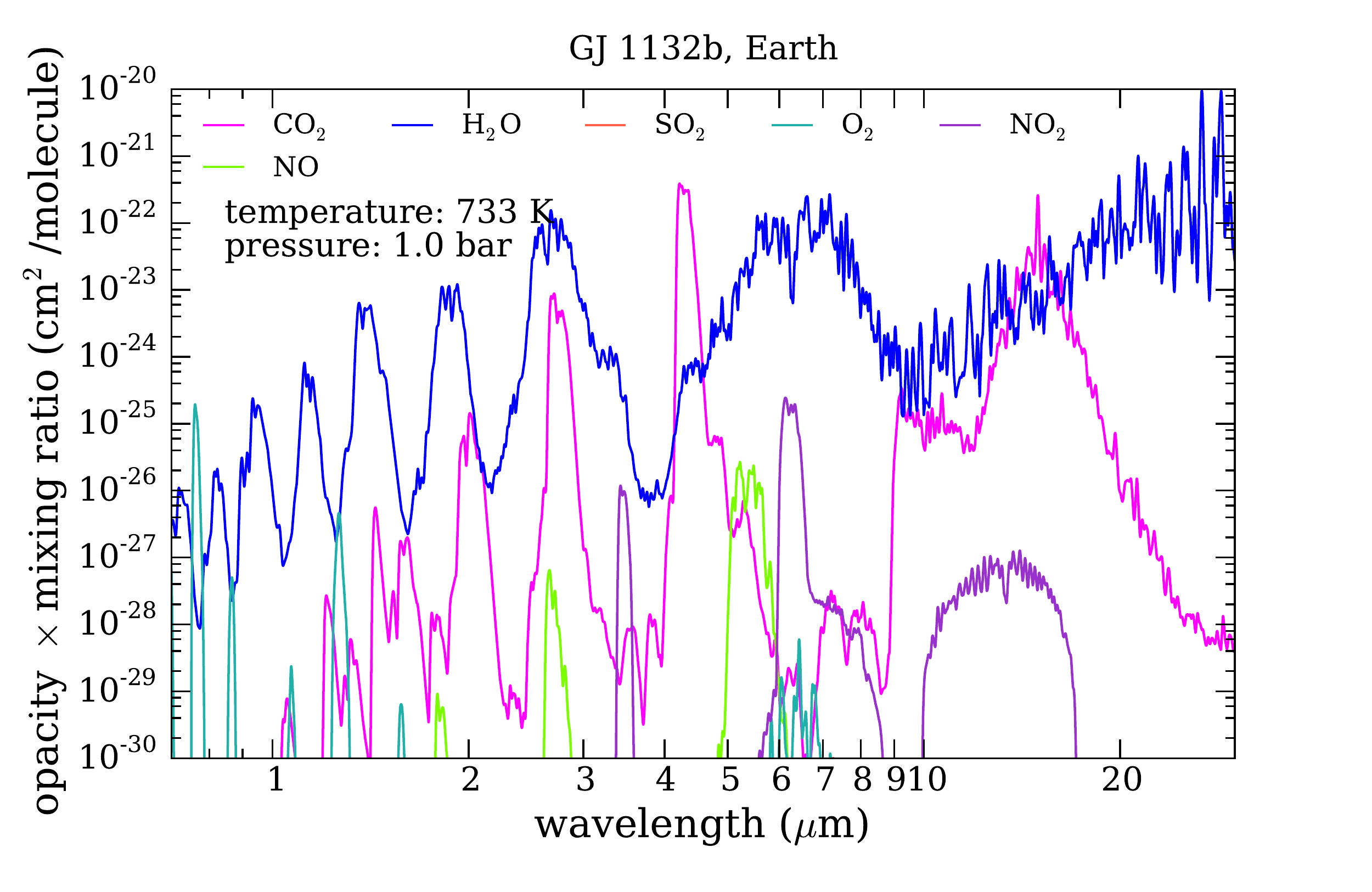}
 \includegraphics[width=3.8in]{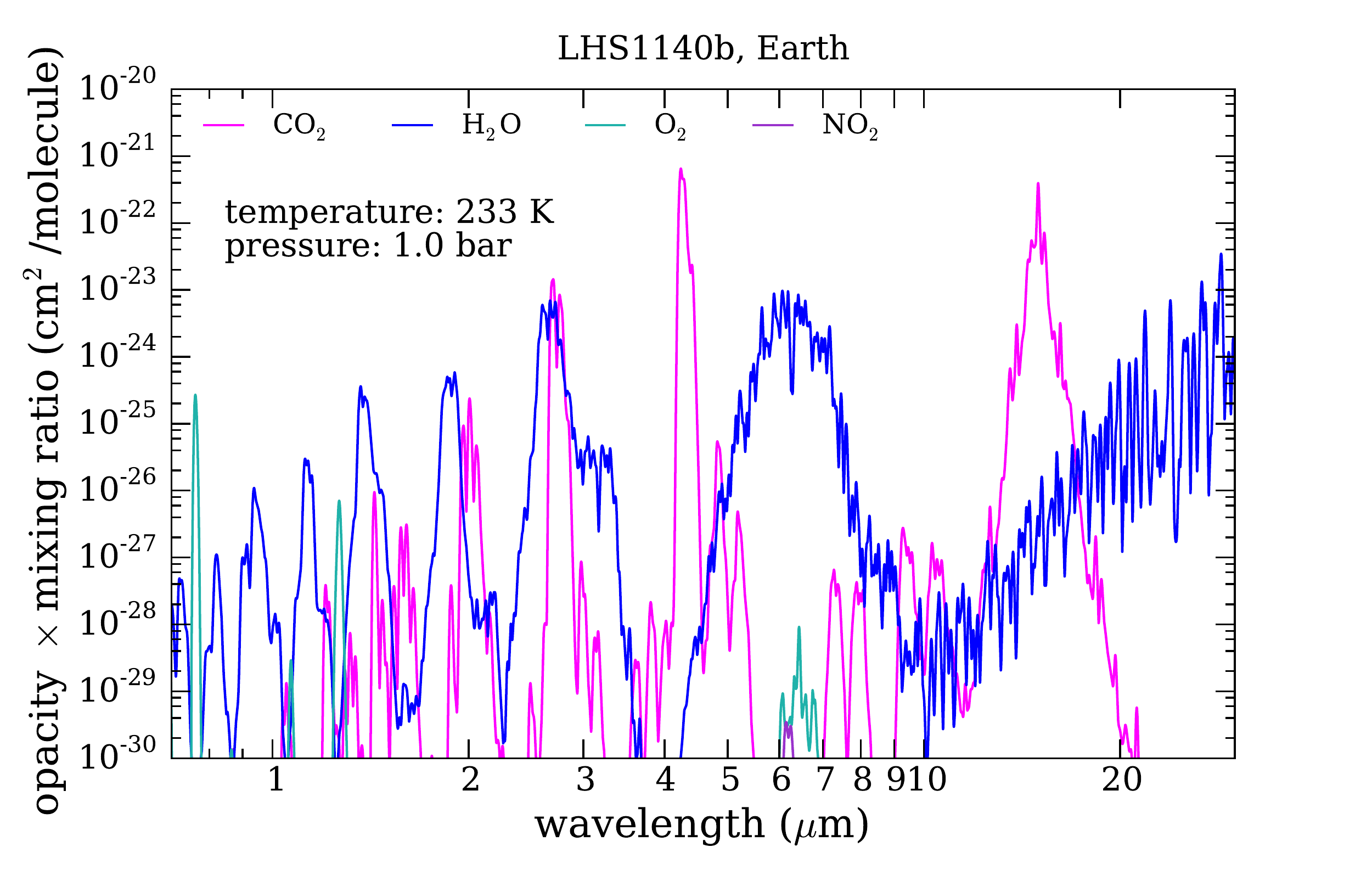}
 \includegraphics[width=3.8in]{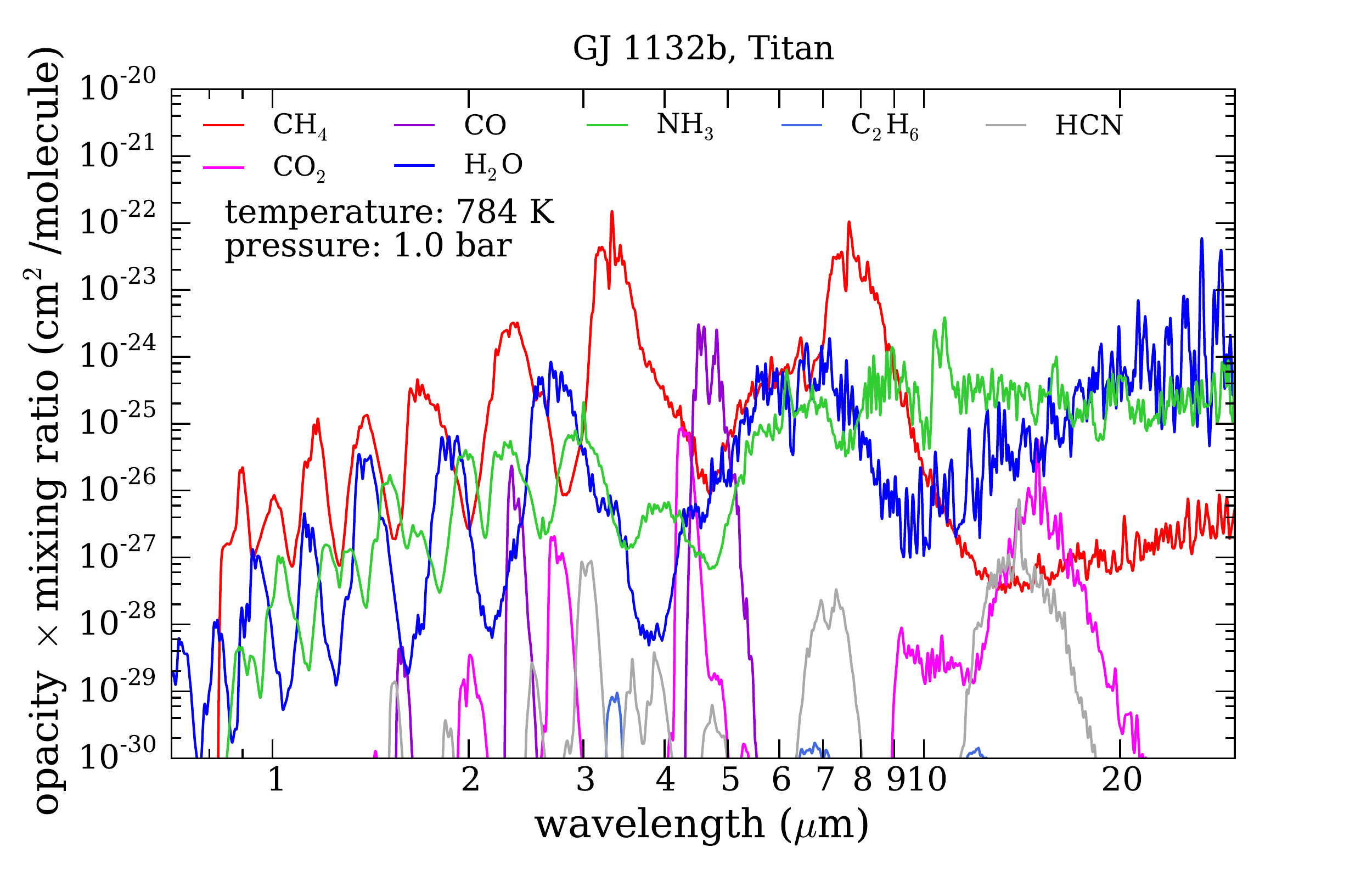}
 \includegraphics[width=3.8in]{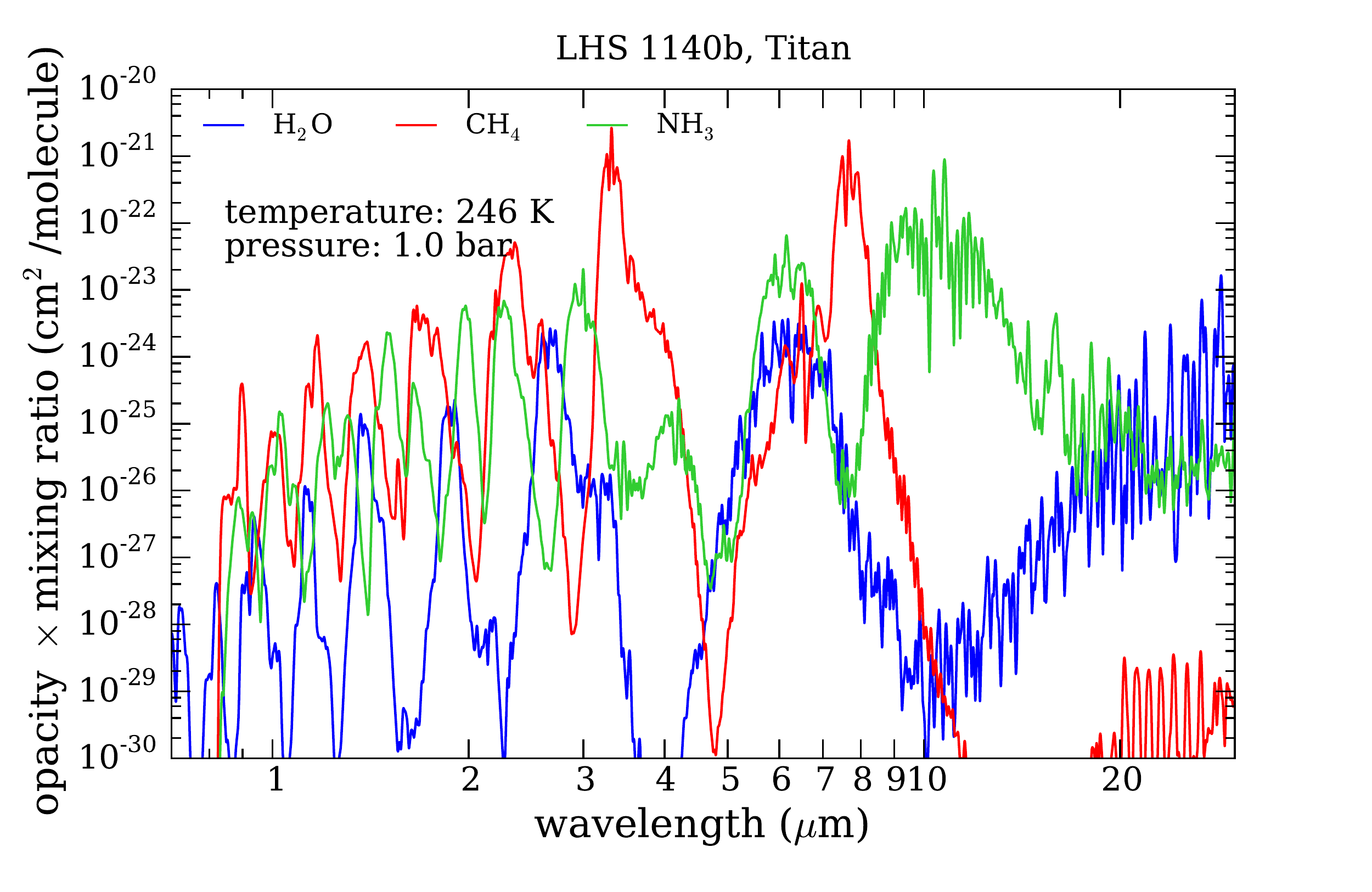}
 \includegraphics[width=3.8in]{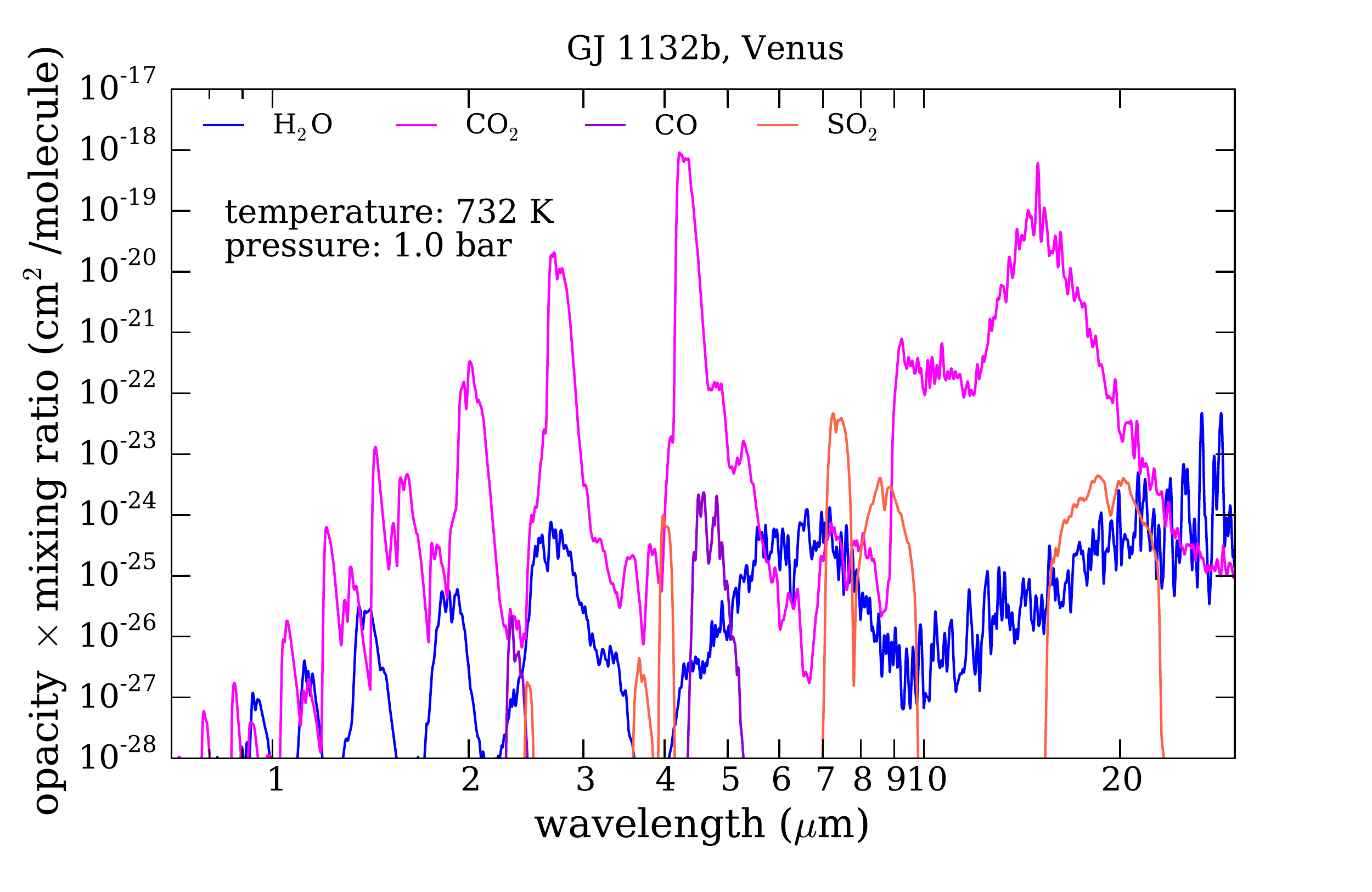}
 \includegraphics[width=3.8in]{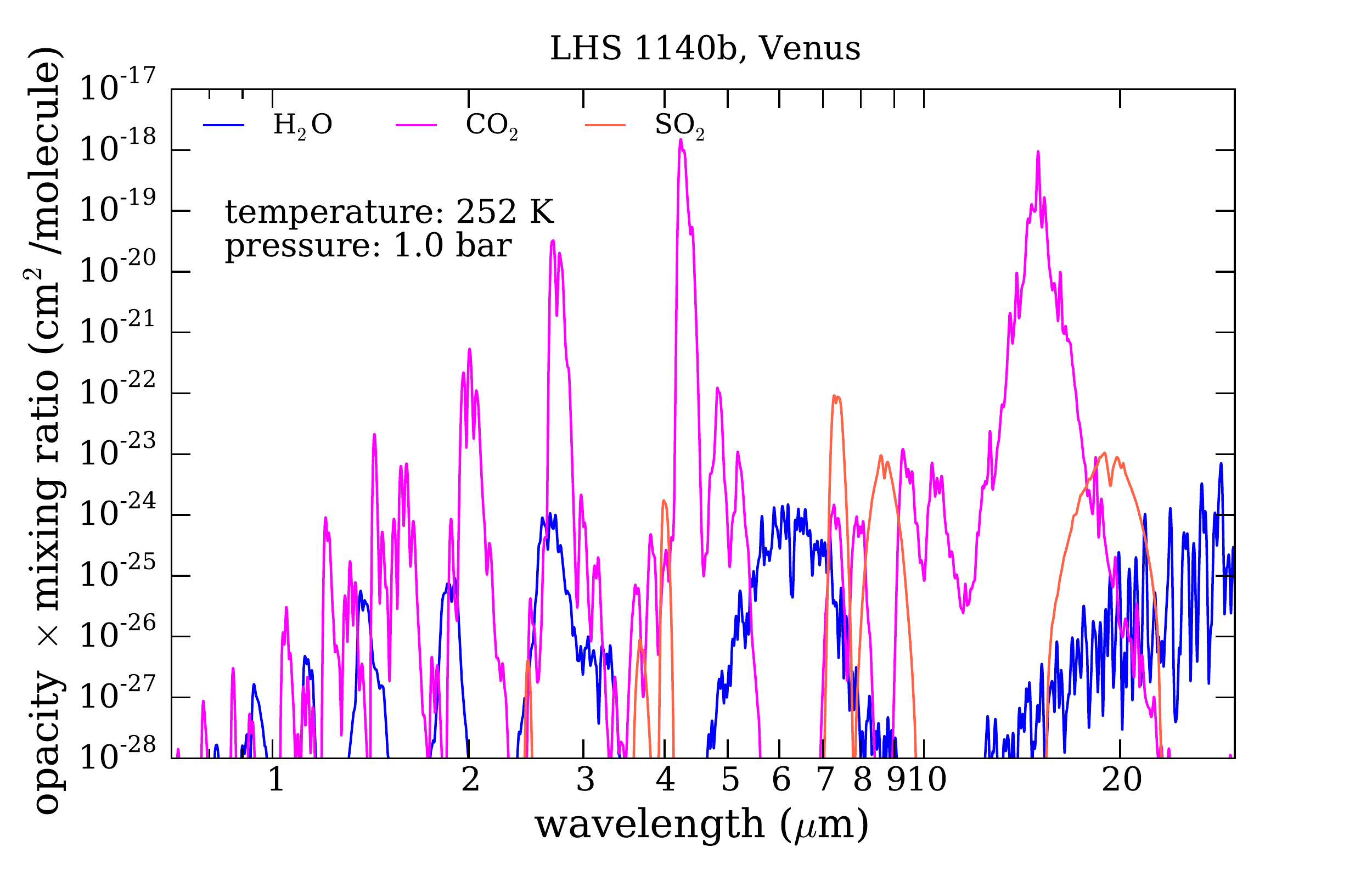}
\caption{Opacities for the major species in six example models are shown. The left column includes models for GJ 1132b and the right column shows those for LHS 1140b. The top row uses Earth-based compositions; the middle row Titan-based compositions, and the bottom Venus-based compositions. For each molecule, the cross section is multiplied by the mixing ratio to show the relative importance of each gas at each wavelength. All cross sections and abundances are calculated for models with 1 bar surface pressure; the surface temperature is indicated at the top left of the plot.}
\label{opacities}
\end{figure*}

Molecular opacities for the three elemental compositions for two example planets are shown in Figure \ref{opacities}. The shapes of the spectral bands are a function of both pressure and temperature and are shown for models with surface pressures of 1 bar (and corresponding surface temperature). The opacities are scaled by the mixing ratio of each species to show the relative importance of each species at a particular wavelength. 

The left column shows the hottest planet in the study, GJ 1132b. The right column shows a colder planet, LHS 1140b. Water vapor is a strong absorber across the infrared for all models, with many rovibrational bands from 0.8 to 20+ $\micron$. For Earth-based compositions (top row), CO$_2$ and O$_2$ also have features at particular wavelengths (notably 4--5 \micron\ and 0.7--0.8 \micron\ respectively). For Titan-based compositions (middle row), CH$_4$ and NH$_3$ are important absorbers for both temperature models, while CO and HCN appear in the hotter model. The Venus-based compositions (bottom row) are dominated by CO$_2$ opacity, with a strong SO$_2$ feature between 7 and 10 \micron.

\subsection{Surface Temperature}

\begin{figure}[h]
\center \includegraphics[width=3.4in]{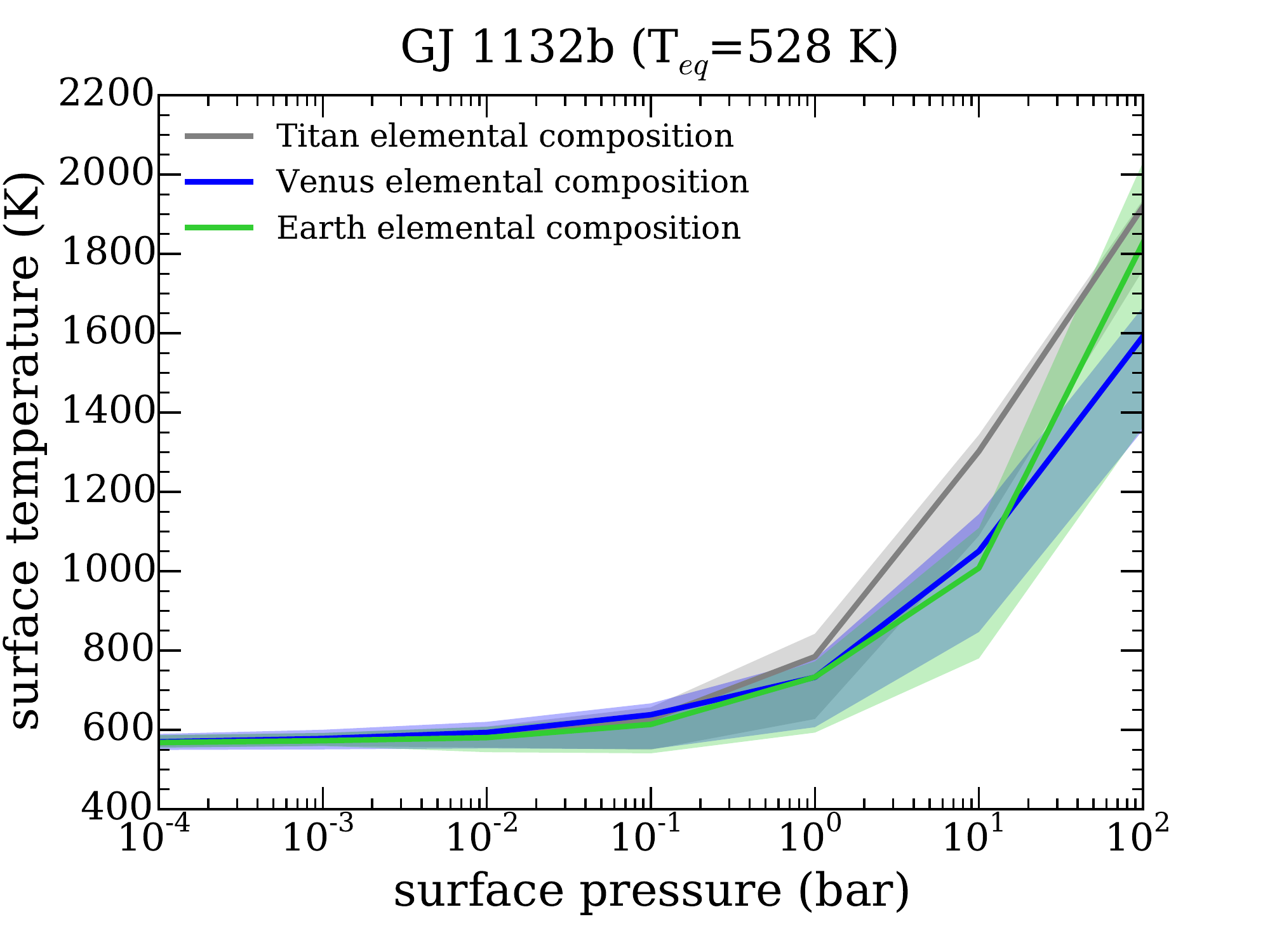}
\vspace{-0.4in}
\center \includegraphics[width=3.4in]{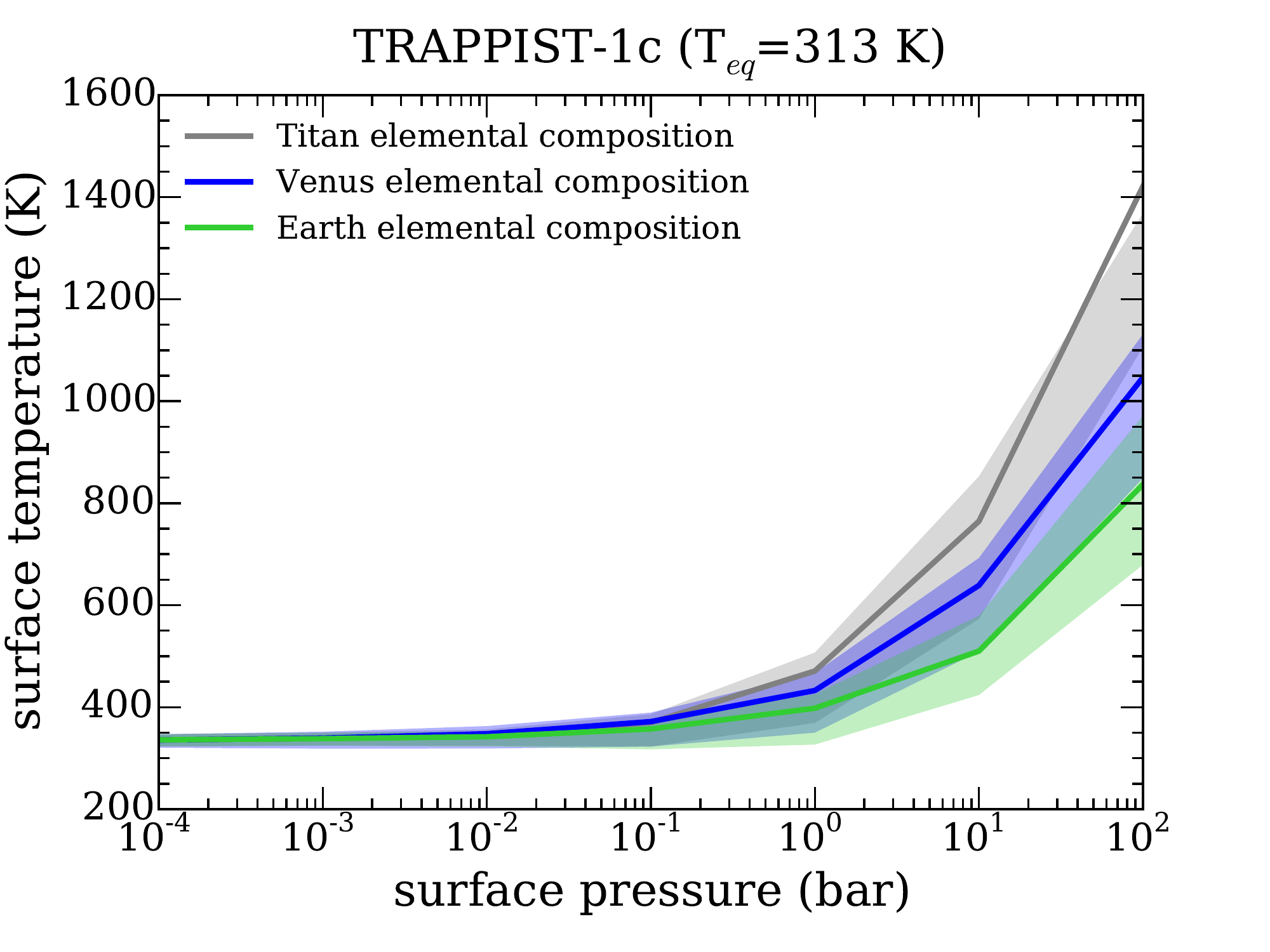}
\vspace{-0.4in}
\center \includegraphics[width=3.4in]{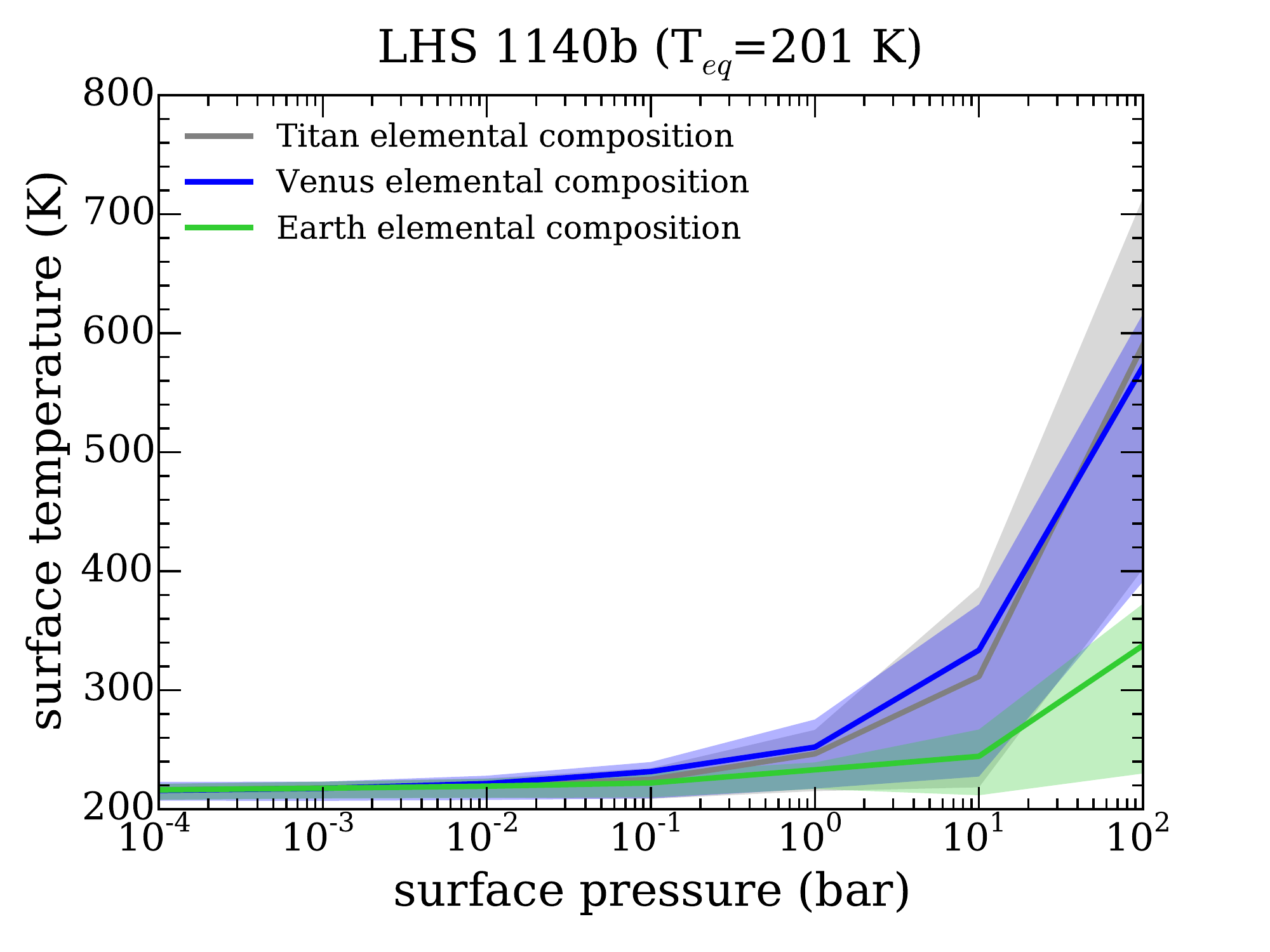}
 \caption{The surface temperature for each planet is shown as a function of surface pressure and atmospheric composition. Three example planets are shown (from top to bottom, GJ 1132b, TRAPPIST-1c, and LHS 1140b. The central darker line represents the surface temperature assuming a Bond albedo of 0.3. The shaded region shows the possible range from highest Bond albedo (0.7) to lowest (0.0). The colors represent the three different elemental compositions studied here. For surface pressures greater than 0.1 bar, the surface is significantly warmer than the equilibrium temperature of the planet. }
\label{psurf_tsurf}
\end{figure}

\begin{figure}[h]
\center \includegraphics[width=3.4in]{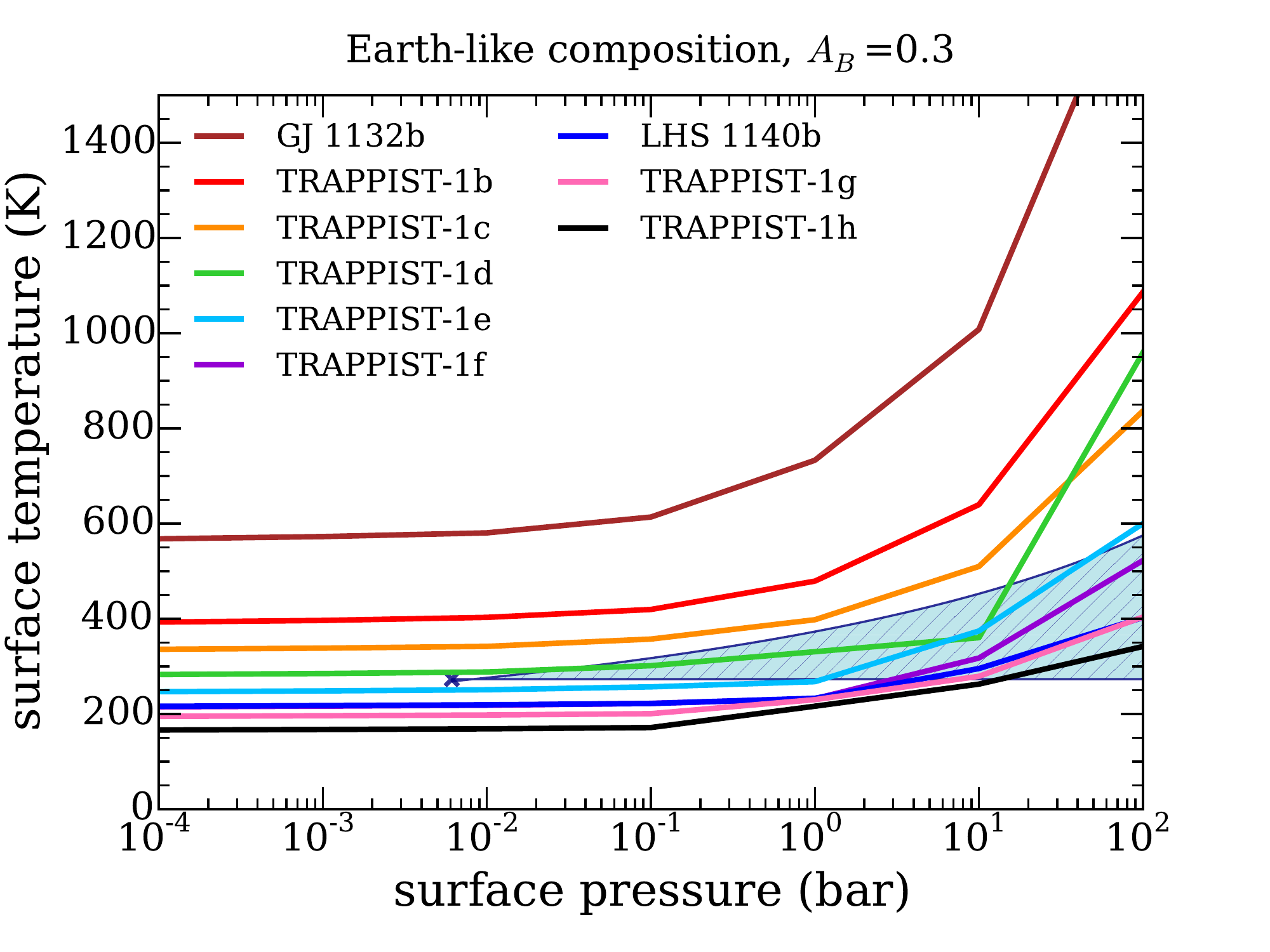}
 \caption{The surface temperature as a function of surface pressure is shown for each of the 9 planets considered here. Each planet is represented by a different colored line; all models have an Earth-based composition and Bond albedo of 0.3. The pressures and temperatures at which water is in the liquid phase is shown as a blue hatched region, as described in the text. The triple point of water is shown as an `x'; at pressures lower than the triple point, water is always either a vapor or a solid. }
\label{all_psurf_tsurf}
\end{figure}

The surface temperature of each model is chosen such that the equilibrium temperature (which depends on the Bond albedo) and effective temperature of the model are equal. Each planet therefore has a range of possible surface temperatures, which depend on the surface pressure of the atmosphere and the composition of gases in that atmosphere. Those ranges are shown in Figure \ref{psurf_tsurf} for three selected planets: GJ 1132b, TRAPPIST-1c, and LHS 1140b. 

As expected, the surface temperature depends most strongly on surface pressure. Higher surface pressures (100 bar) similar to Venus's atmosphere result in surface temperatures that are $\sim$3--5$\times$ higher than the equilibrium temperature. Atmospheres more like Earth's (1 bar) have a less strong warming of the surface, with surface temperatures around 1.1--1.6$\times$ the equilibrium temperature. 

The Bond albedo ($A_B$) of the planet also strongly affects the surface temperature. The Bond albedo is the ratio of the energy reflected by a planet to the energy incident upon the planet. It is a function of both the reflectivity of a planet as a function of wavelength and the host stellar spectrum. In general, planets around later type stars will have lower Bond albedos than around earlier type stars even if the geometric albedo of the planet is identical in both cases \citep[see, e.g.,][]{Selsis07}. As a point of reference, we ran a simple case to test the dependence of the spectral type of the star on the Bond albedo for a Venus-like model \citep[][Robinson \& Crisp, submitted]{Meadows96}. We found that the Bond albedo of the model Venus around a late M dwarf was about 0.1 lower than around a G dwarf like the Sun (0.7 for the M dwarf and 0.8 for the G dwarf). 

For simplicity we take Bond albedos of 0.7 and 0.0 as conservative upper and lower bounds, giving us a wide range in possible surface temperatures for each surface pressure and composition. For example, for a 100 bar, Venus-composition atmosphere around TRAPPIST-1c, we find a surface temperature range from 850 K ($A_B$=0.7) to 1050 ($A_B$=0.3), to 1125 ($A_B$=0.0). 

Lastly, the composition affects the surface temperature. This is most evident in the bottom panel of the Figure \ref{psurf_tsurf}, showing LHS 1140b which has an equilibrium temperature of $\sim$200 K. For this temperature, much of the atmosphere is cold enough that water condenses, removing it from the atmosphere. Water is the most important greenhouse gas in an Earth-based atmosphere. For Venus- and Titan-based compositions, other greenhouse gases such as carbon dioxide and methane remain in high abundance. However, for the Earth-based composition at this temperature, with a dominant absorber removed from much of the atmosphere, the Earth-based models have lower surface temperatures than the other two compositions. 

Figure \ref{all_psurf_tsurf} shows the surface temperature as a function of surface pressure for all nine planets studied here, assuming an Earth-based composition and Bond albedo (0.3) (solid green lines from Figures \ref{psurf_tsurf}). For illustration, the temperature at which water can be liquid is shown as the blue hatched region, calculated using a constant melting point across this pressure region (273 K) and a boiling point calculated using the Clausius-Clapeyron relationship, 

\begin{equation}
    T_B = \left( \dfrac{1}{T_0} - \dfrac{R \ln \frac{P}{P_0}}{\Delta H_{vap}} \right) ^{-1}
\end{equation}
where $T_0=373.15$ is the boiling point of H$_2$O at 1 bar, $R$ is the gas constant 8.314 J mol$^{-1}$K$^{-1}$, $P$ is the pressure, $P_0$ is 1 bar, and $\Delta H_{vap}$=40660 J mol$^{-1}$ is the heat of vaporization for H$_2$O. 

For this particular set of assumptions, TRAPPIST-1d has a surface temperature amenable to surface liquid water at 0.1–10 bar; TRAPPIST-1e does with a surface pressure of 10 bar; and TRAPPIST-1f, TRAPPIST-1g, and LHS 1140b do with surface pressures of 100 bar. Note that we do not calculate model P--T profiles in radiative-convective equilibrium or take into consideration climactic effects such as ice-albedo feedback, so these models are meant to be illustrative of a broad range of atmospheres, not definitive estimates of liquid-water habitable zones, which have previously been published in many other relevant works \citep[see, e.g.,][]{Shields16, Kasting93, Kopparapu13, Selsis07, Wordsworth11}. 

\subsection{Thermal Emission Spectra}  

\begin{figure}[tbh]
\center \includegraphics[width=3.6in]{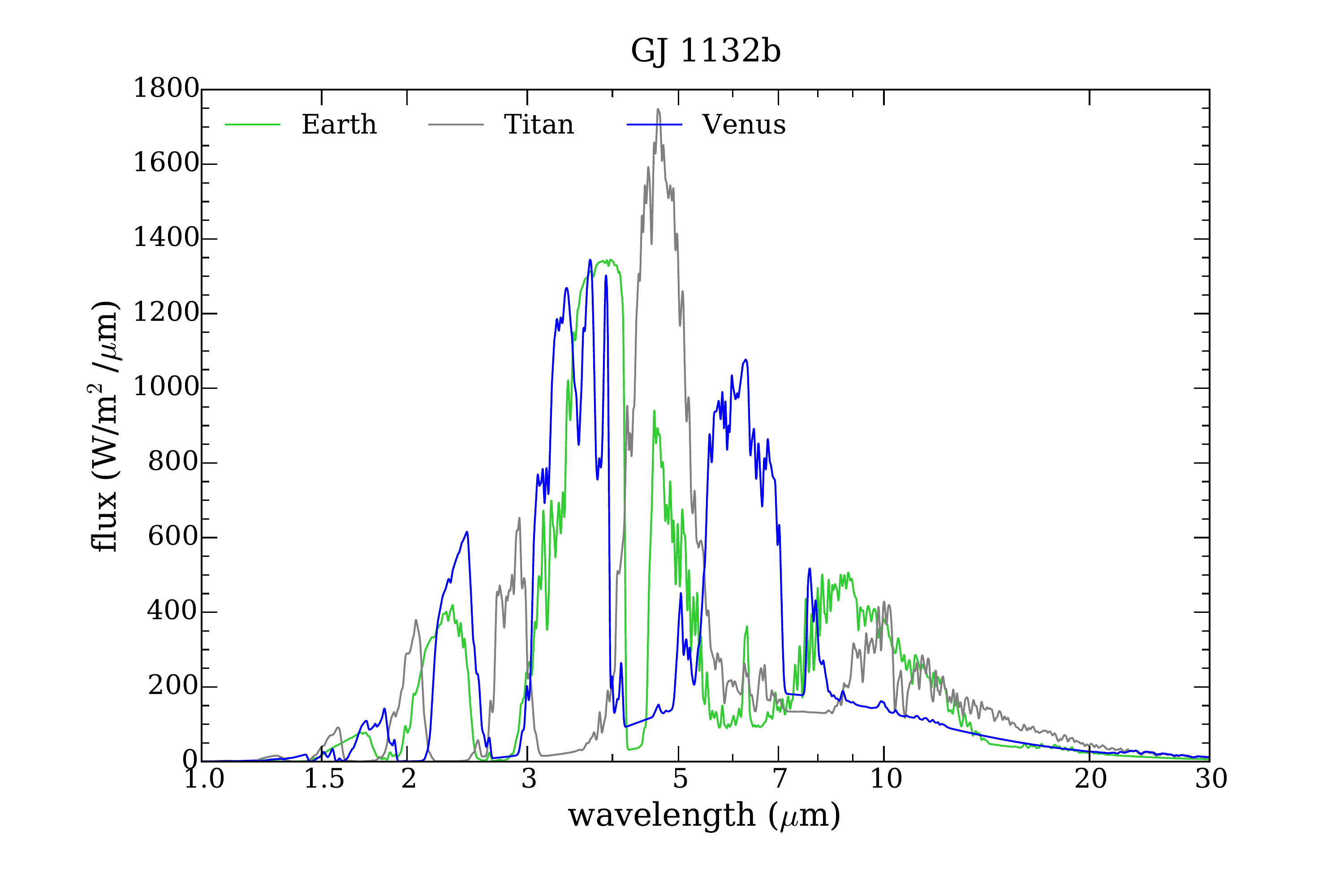}
\vspace{-0.4in}
\center \includegraphics[width=3.6in]{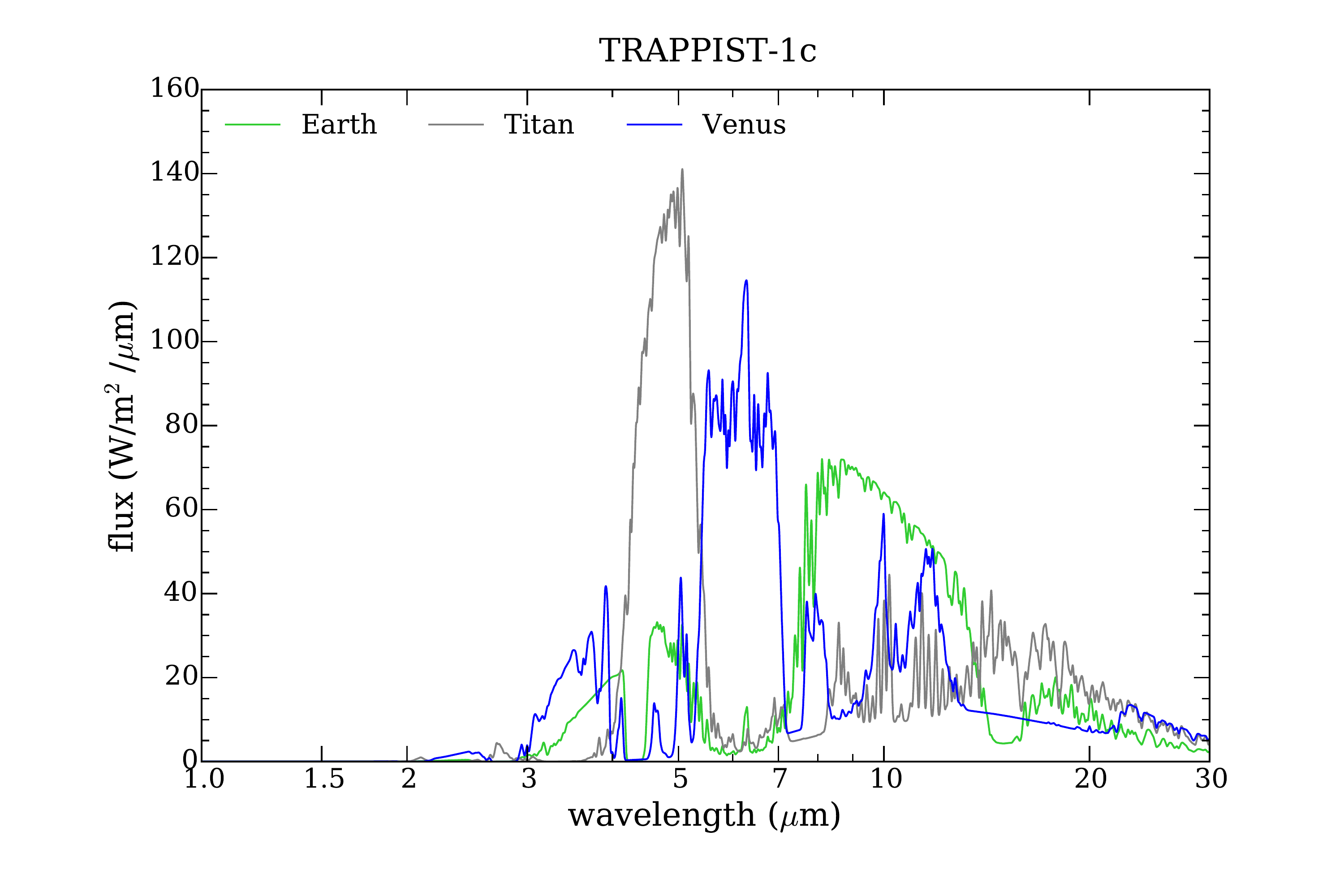}
\vspace{-0.4in}
\center \includegraphics[width=3.6in]{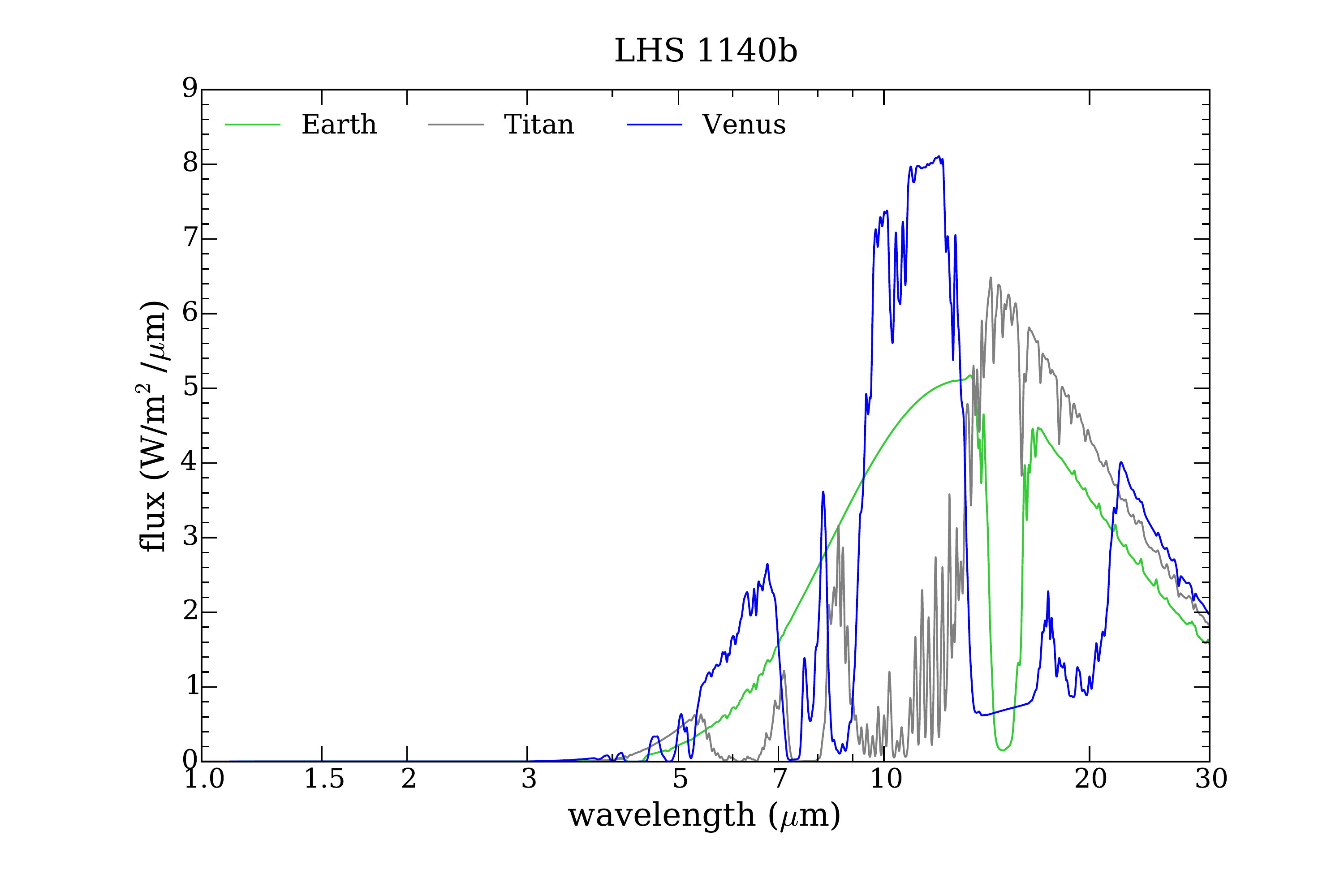}
 \caption{Thermal emission spectra for three example planets with 3 compositions are shown. From top to bottom, we show spectra from the warmest to coolest planet: GJ 1132b, TRAPPIST-1c, and LHS 1140b. The different colored lines represent different elemental compositions (Venus, Earth, and Titan). Each model assumes a surface pressure of 1 bar. Note that both the composition and the equilibrium temperature strongly control the spectrum. }
\label{spectra_3planets}
\end{figure}

\begin{figure}[tbh]
\hspace{-0.9cm}
 \includegraphics[width=4.0in]{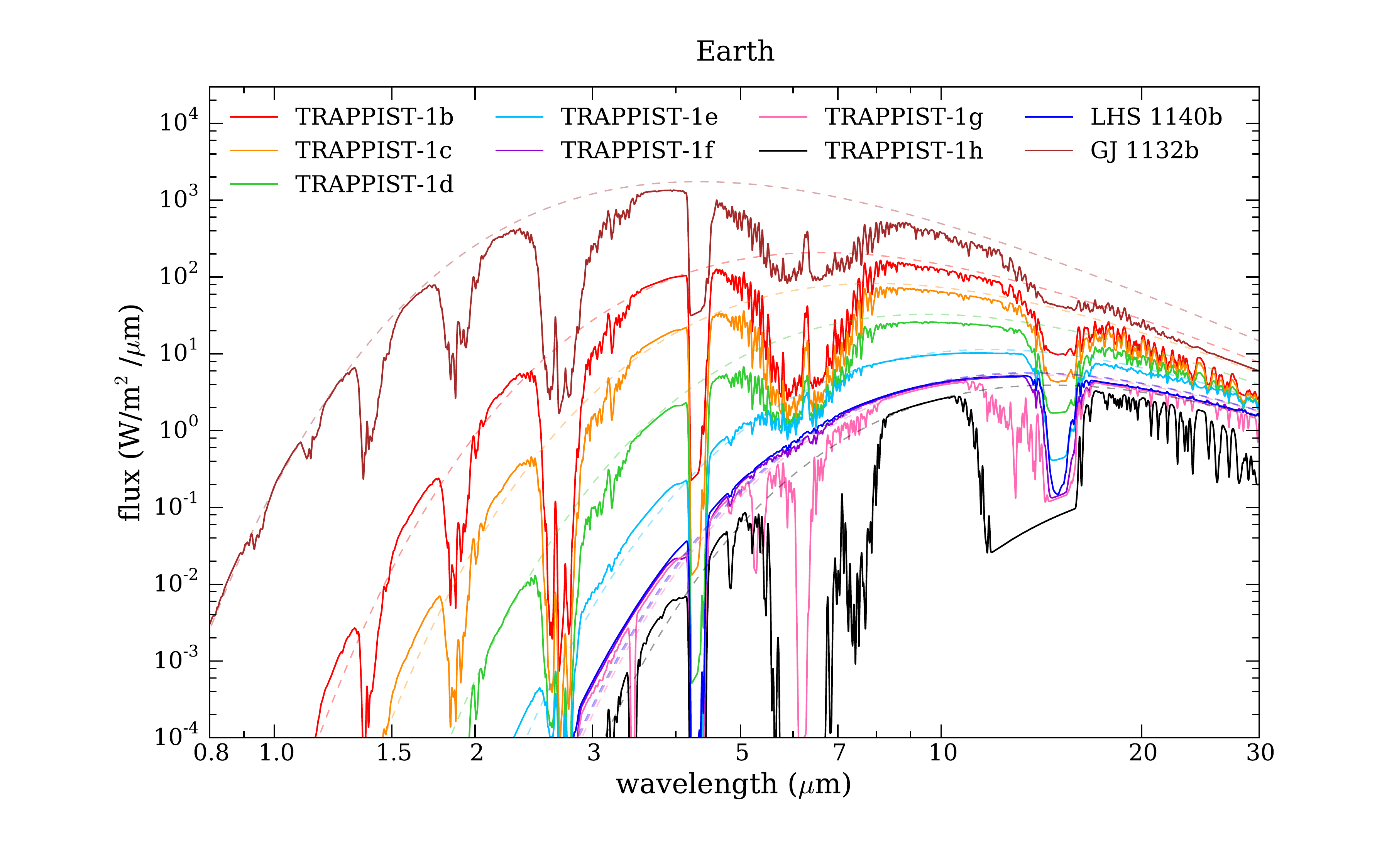}
 \caption{Thermal emission spectra for each planet, assuming a surface pressure of 1 bar and Earth-based composition. Approximate emission from the surface of the planet is represented by the dashed lines, which are blackbody curves for each planet's surface. }
\label{spectra_allplanets}
\end{figure}

\begin{figure}[tbh]
\includegraphics[width=3.7in]{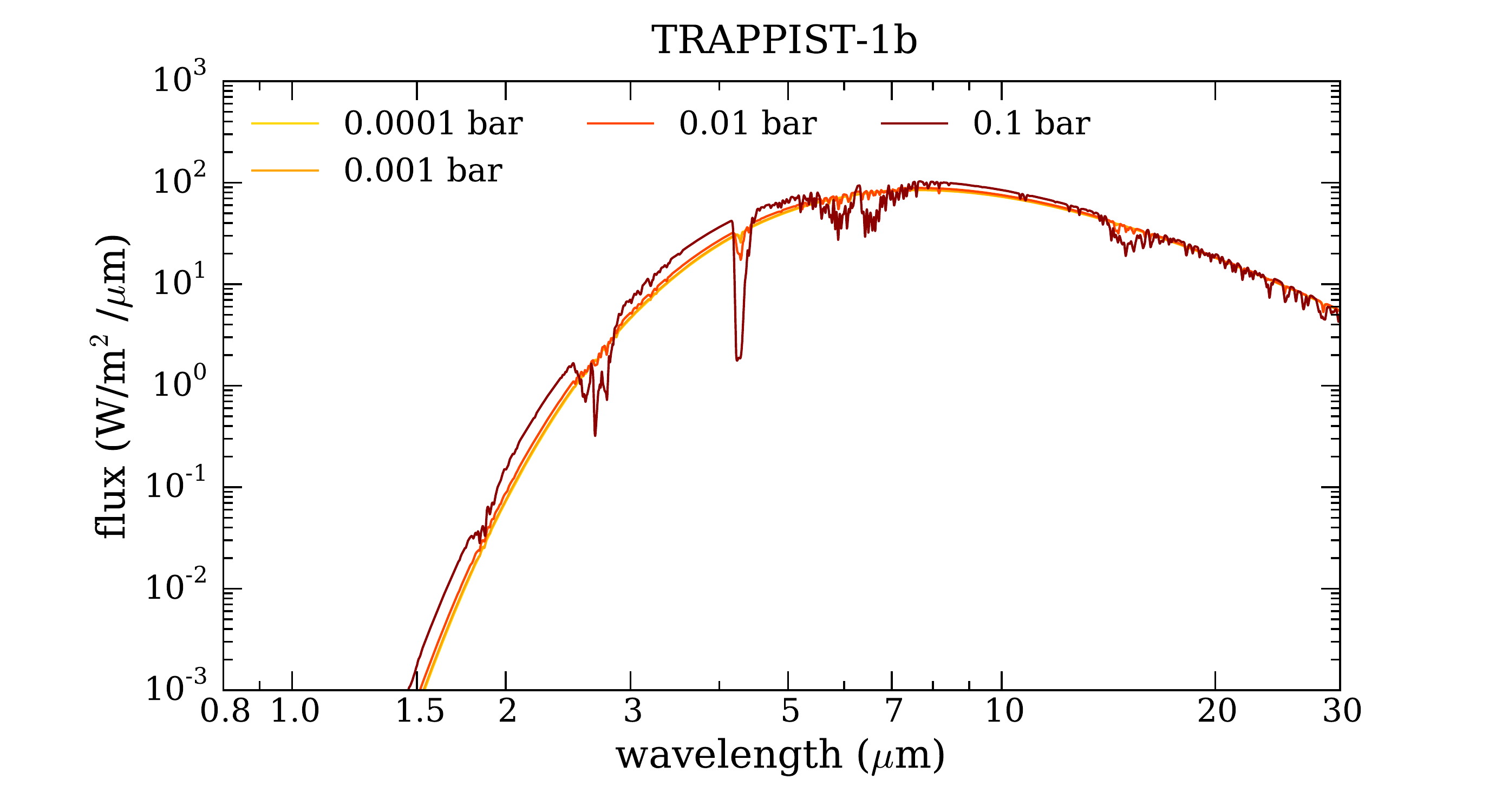}
 \includegraphics[width=3.7in]{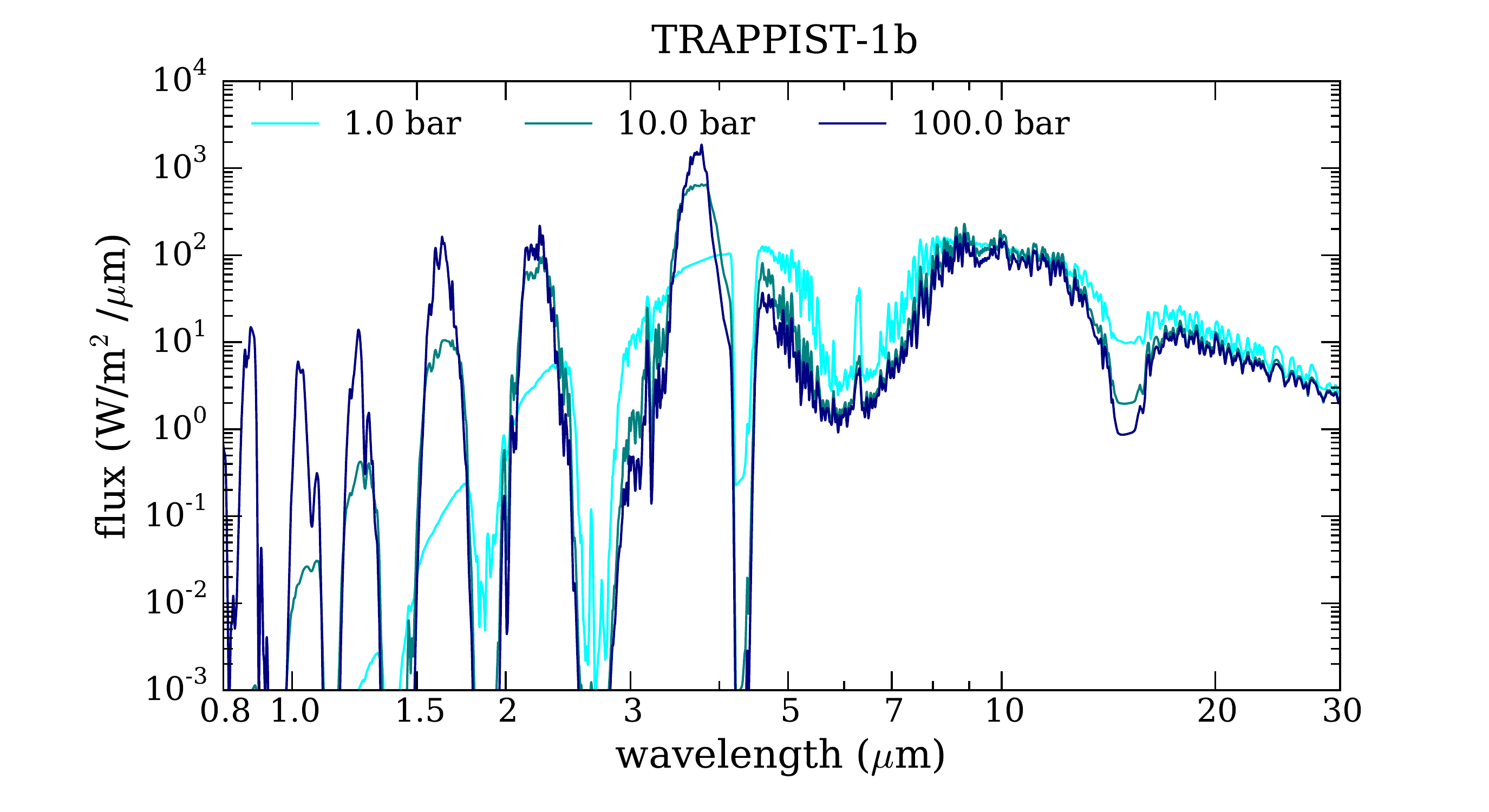} 

 \caption{Thermal emission spectra for TRAPPIST-1b for each surface pressure (0.0001 bar to 100 bar). All models assume Earth-based composition. The top panel includes thinner atmospheres ($\leq0.1$ bar) and the bottom panel includes thicker atmospheres ($\geq$1 bar). For low surface pressures, the thermal emission is blackbody-like, because the optical depth of the atmosphere is low and the spectrum is dominated by the surface. For higher pressures, the spectrum has deep absorption features, with regions of high flux where the hot deep layers of the atmosphere are probed. }
\label{spectra_pressures}
\end{figure}

The model grid of thermal emission spectra includes 384 model spectra (9 planets, 7 surface pressures, 3 compositions, and 2 albedos (0.0 and 0.3). Examples of these thermal emission spectra that illustrate the effect of each parameter are shown in Figures \ref{spectra_3planets}, \ref{spectra_allplanets}, and \ref{spectra_pressures}. For each of these figures, the emergent flux at the top of the planet's atmosphere is shown. To calculate eclipse depths, the surface flux must be multiplied by the surface area of the planet and divided by the total flux from the star: 

\begin{equation}
\frac{F_{surf,p} R_p^2} {F_{surf,s} R_s^2},
\end{equation}
where $F_{surf,p}$ is the surface flux from the planet, $R_p^2$ is the planet radius, $F_{surf,s}$ is the surface flux from the star, and $R_s^2$ is the stellar radius. 

Examples of thermal emission spectra for three example planets (GJ 1132b, TRAPPIST-1c, and LHS 1140b) are shown in Figure \ref{spectra_3planets}. These examples all assume surface pressures of 1 bar and Bond albedos of 0.3. For each planet, the three elemental compositions are shown. The composition has a strong effect on the emergent spectra, with different molecules carving the spectrum. The temperature also plays a major role, controlling both the abundances of molecules in chemical equilibrium as well as the dominant wavelengths of thermal emission; the wavelengths of peak thermal emission shift from $\sim$2--10 \micron\ for the warmest target, GJ 1132b, to $\sim$3--20 \micron\ for TRAPPIST-1c, to $\sim$4--30 \micron\ for LHS 1140b. 

An example spectrum for each planet is shown in Figure \ref{spectra_allplanets}, assuming an Earth-based composition and 1 bar surface pressures. The dashed lines show the approximate emission from the planetary surface. In some wavelength regions (for example, 8--12 \micron\ for the 4 coldest planets), the spectrum follows the surface emission, indicating that we are seeing down to the planet's surface at those wavelengths. Other regions that look smooth like blackbodies but are significantly cooler than the surface are from regions we probe the isothermal regions of the atmosphere.

The effect of surface pressure on predicted thermal emission is substantial, as shown in Figure \ref{spectra_pressures}. For thin atmospheres ($\leq0.01$ bar), the thermal emission is dominated by emission from the planet's surface and is therefore blackbody-like, assuming a uniformly emissive surface. As the surface pressure increases, the surface temperature increases; these deep layers are visible in regions with low molecular absorption. This effect means that, in general, models with higher surface pressures will result in emission spectra with increased flux at short wavelengths and strong absorption features.

\subsection{Transmission Spectra} 

\begin{figure*}[tbh]
\center \includegraphics[width=6.0in]{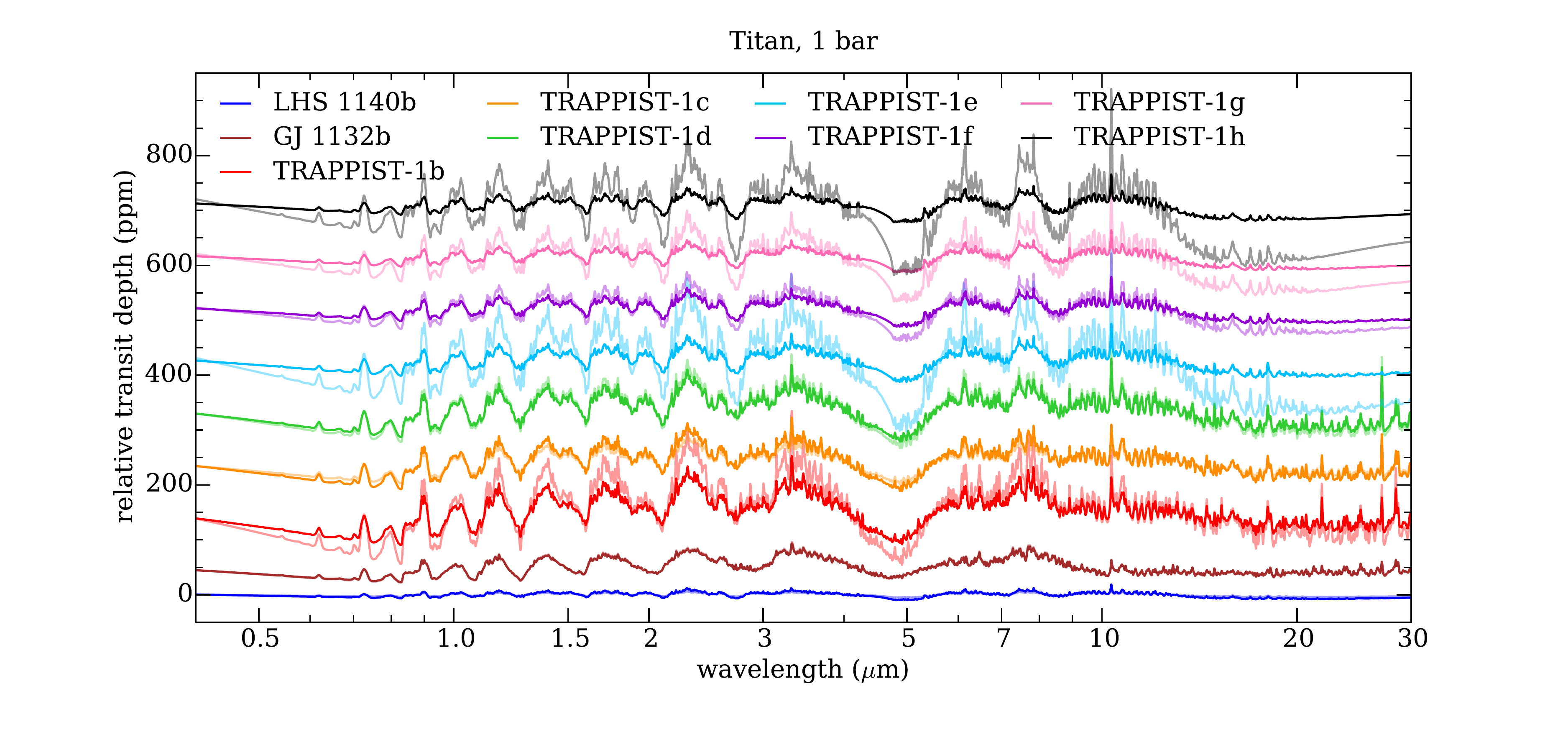}
\vspace{-0.1in}
\center \includegraphics[width=6.0in]{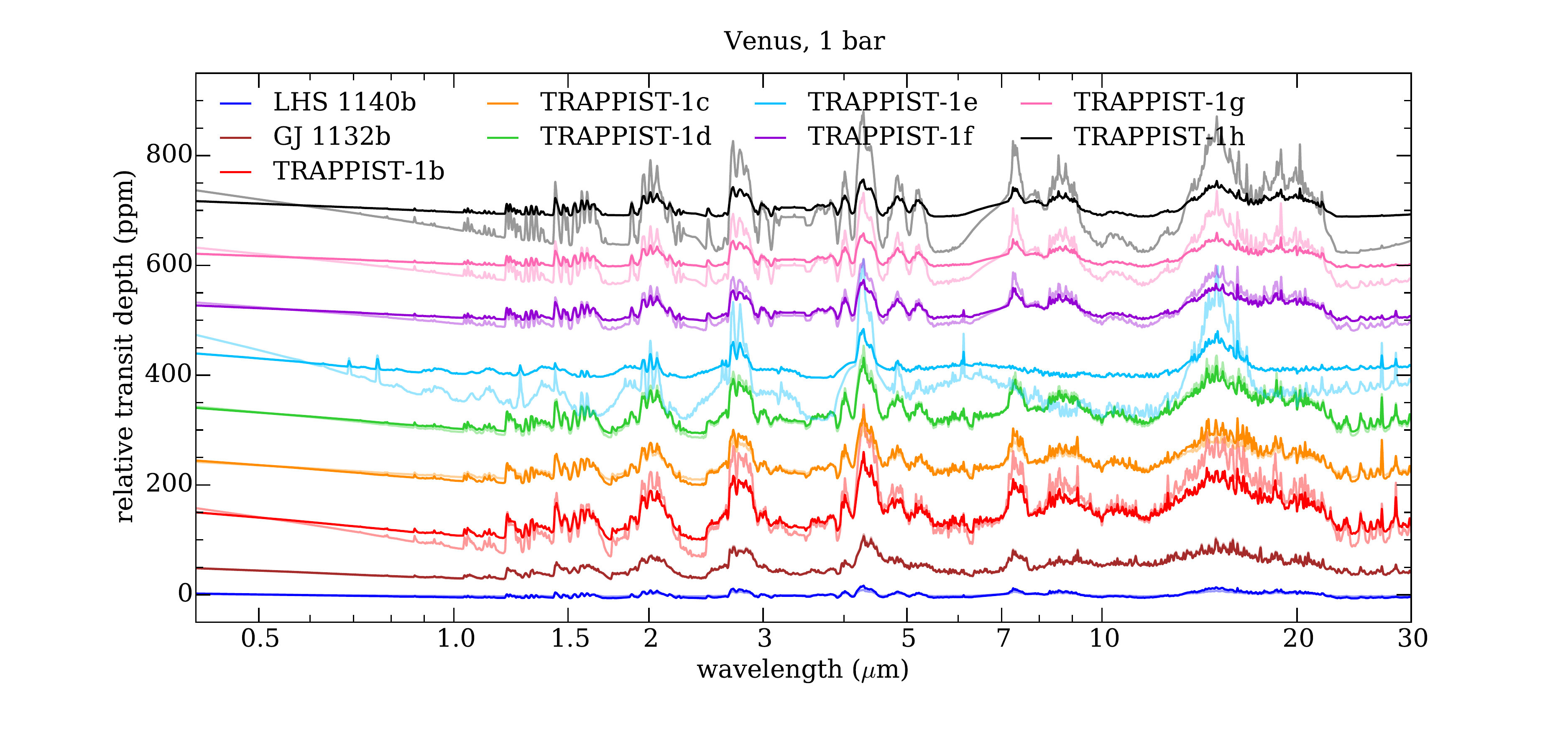}
\vspace{-0.1in}
\center \includegraphics[width=6.0in]{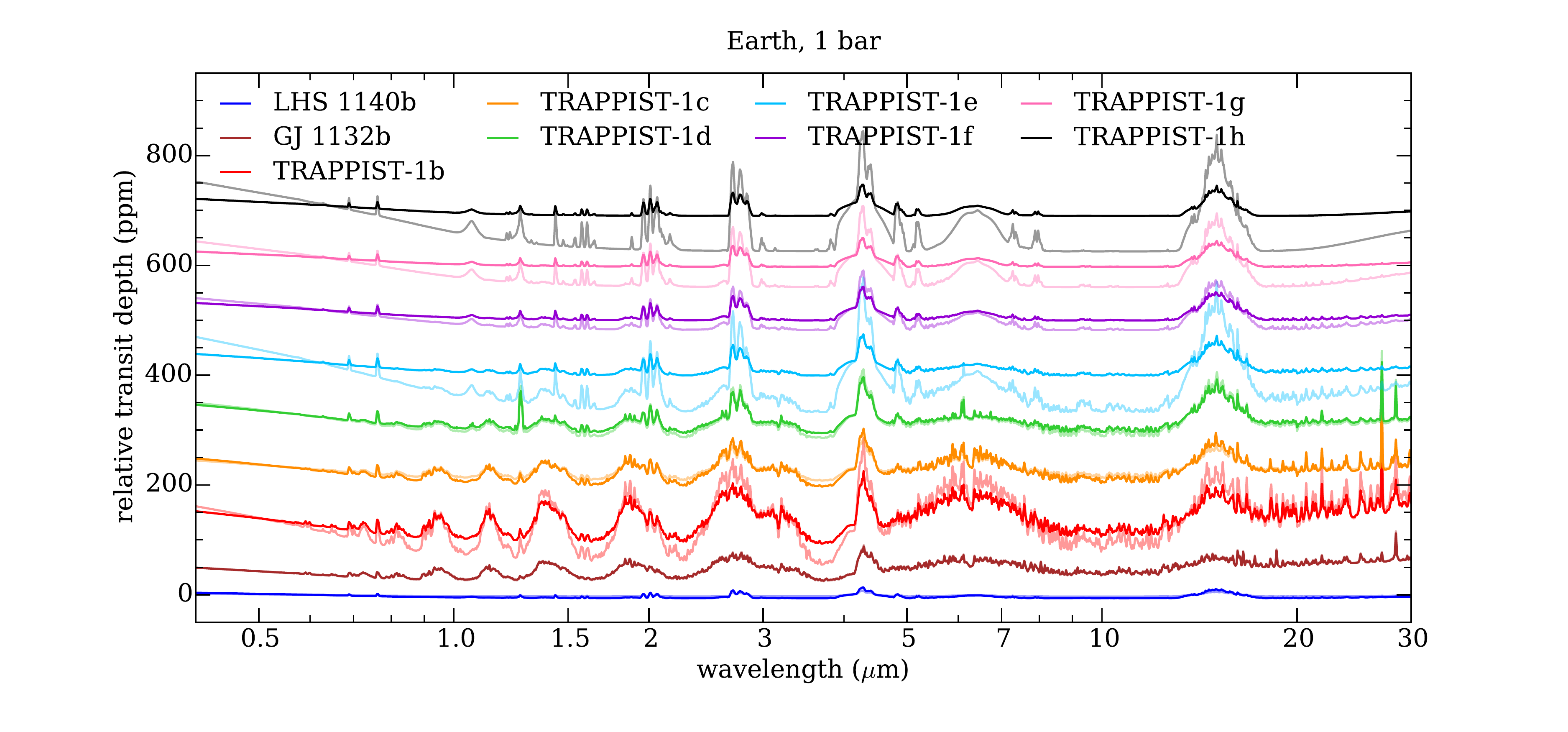}
 \caption{Transmission spectra for each planet are shown in each panel. Solid lines show models that are calculated using a surface gravity corresponding to an Earth-like rock/iron composition, while the transparent lines show models that assume a gravity corresponding to the measured mass. The surface pressure for all models is 1.0 bar. From top panel to bottom panel, the compositions are Titan-based, Venus-based, and Earth-based. Vertical offsets between transmission spectra are added for clarity.} 
\label{transspectra_allplanets}
\end{figure*}

\begin{figure*}[tbh]
\center \includegraphics[width=5.9in]{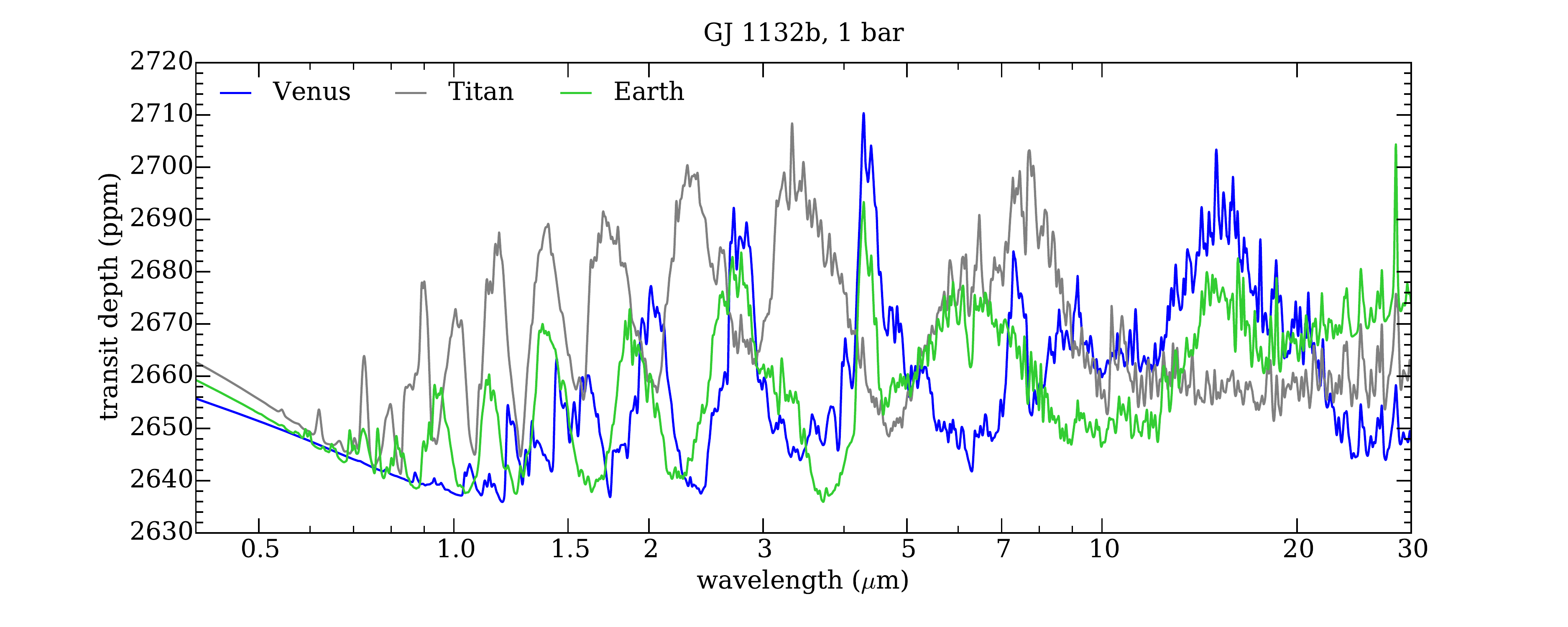}
\vspace{-0.13in}
\center \includegraphics[width=5.9in]{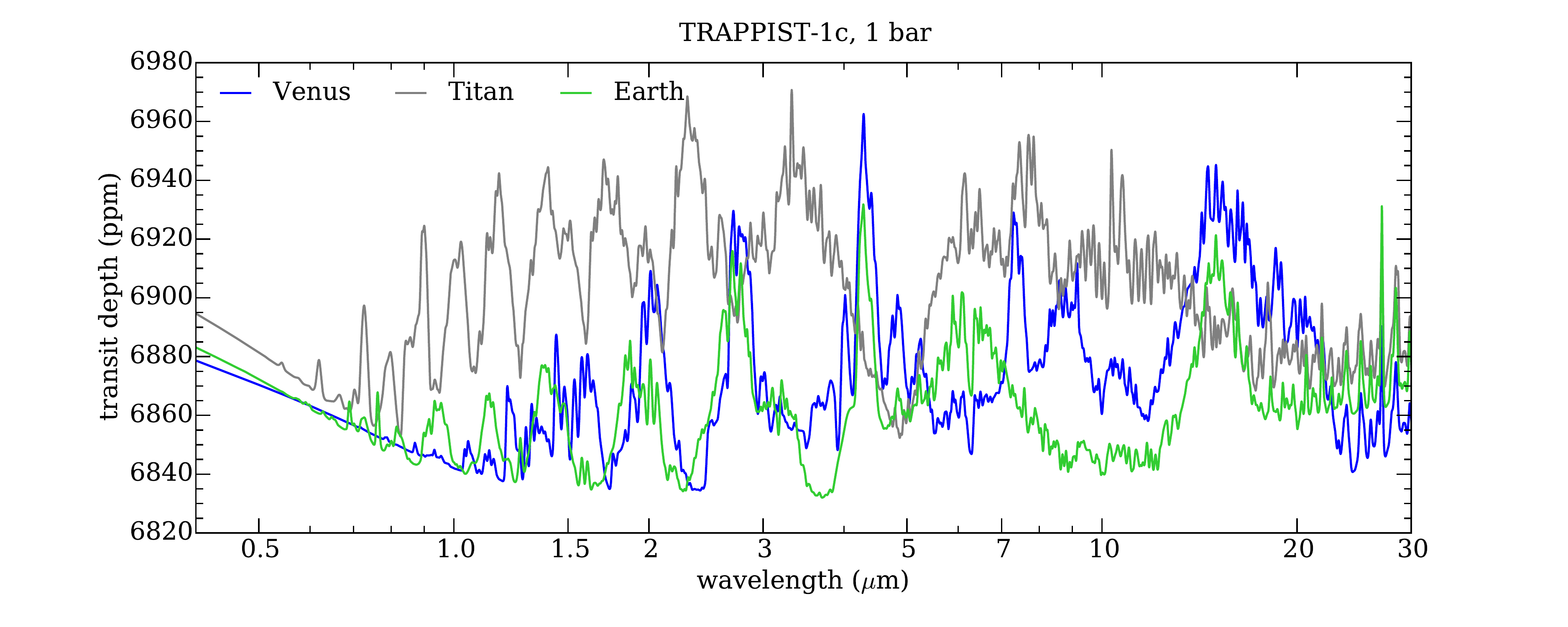}
\vspace{-0.13in}
\center \includegraphics[width=5.9in]{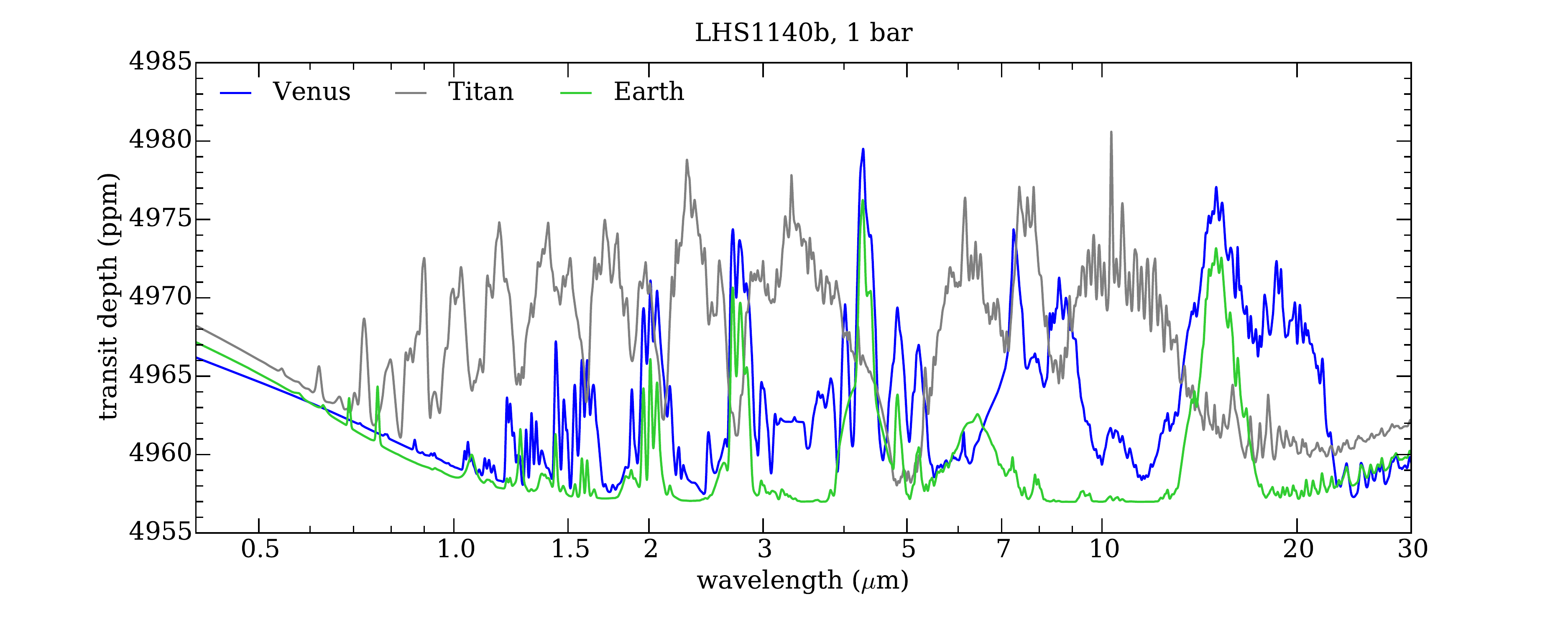}
 \caption{Transmission spectra for three example planets: GJ 1132b (top), TRAPPIST-1c (center), and LHS 1140b (bottom). In each panel, models with each of the three compositions (Earth-, Venus-, and Titan-based) are shown. All models assume a surface pressure of 1 bar. Note that the composition strongly affects the locations of features in the transmission spectra.}
\label{transspectra_3planets}
\end{figure*}

\begin{figure}[tbh]
\center \includegraphics[width=3.5in]{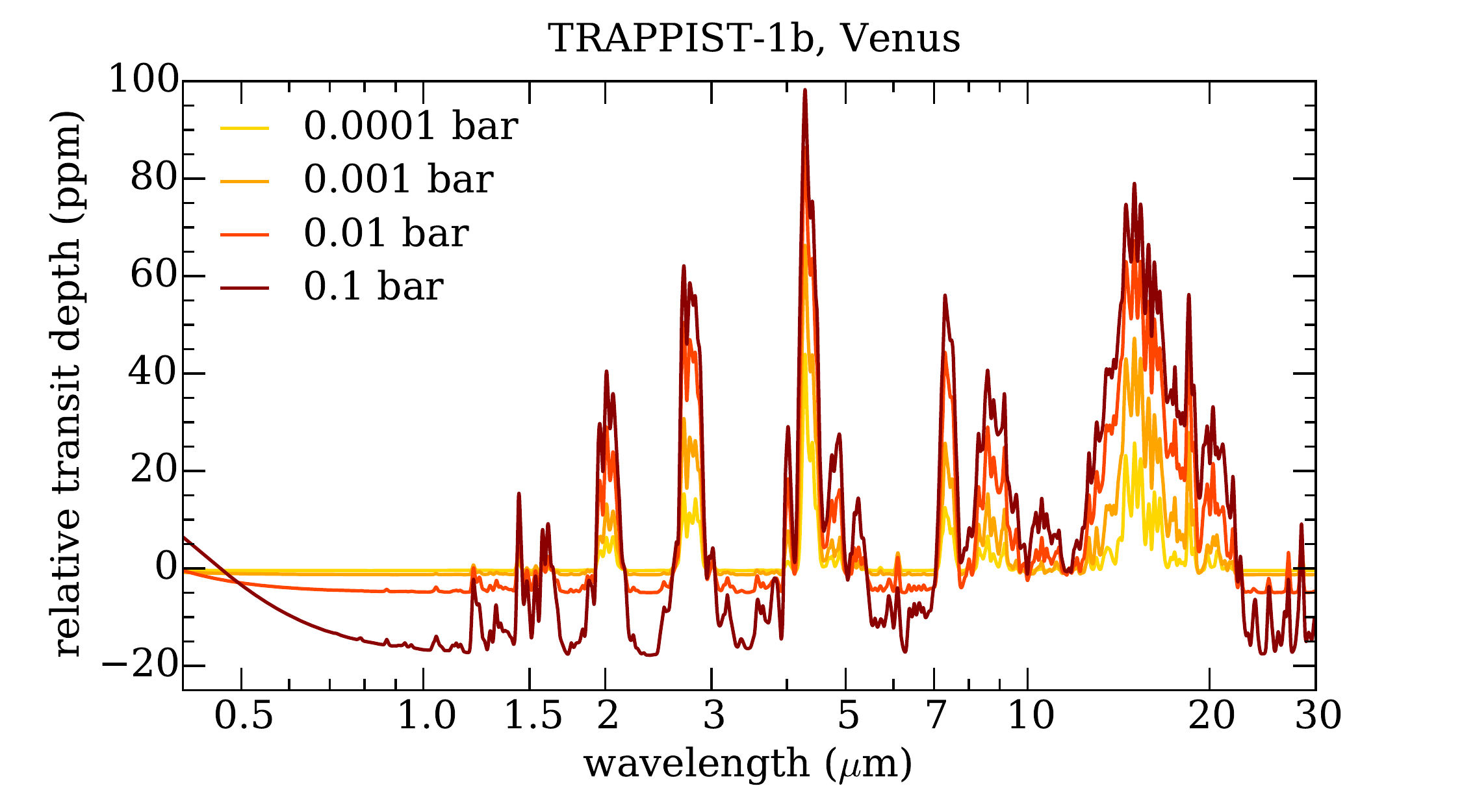}
 \center \includegraphics[width=3.5in]{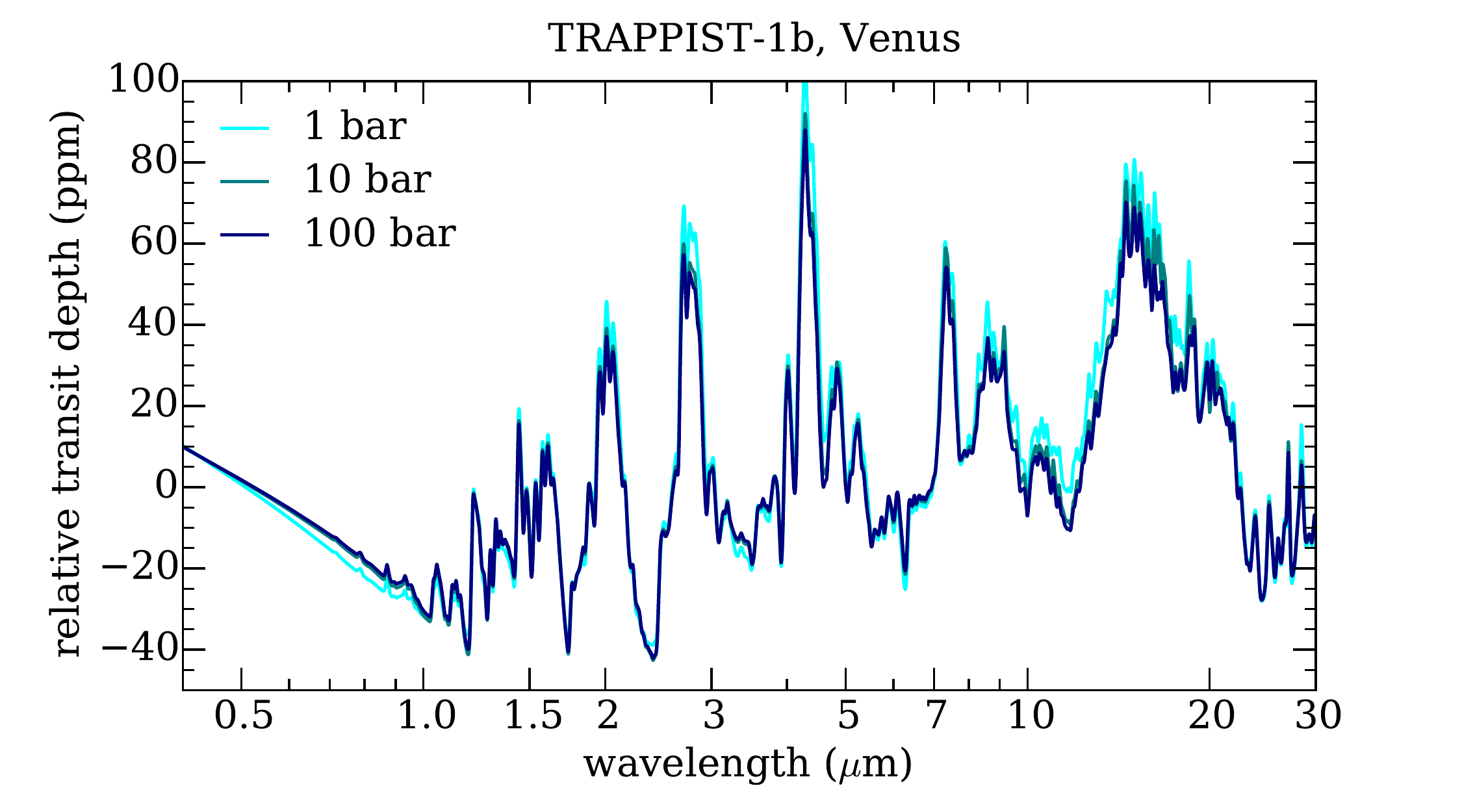}
\caption{Transmission spectra for TRAPPIST-1b for each surface pressure (0.0001 bar to 100 bar). All models assume Venus-based composition. The top panel includes thinner atmospheres ($\leq$0.1 bar) and the bottom panel includes thicker atmospheres ($\geq$1 bar).  For very low surface pressures, the amplitude of features is small and the spectrum has flat sections, where the surface is probed. The amplitude of features increases with increasing surface pressure. However, for higher surface pressures the amplitude of features is approximately the same regardless of surface pressure. }
\label{transspectra_pressure}
\end{figure}

The model grid of transmission spectra includes 384 model spectra (9 planets, 7 surface pressures, 3 compositions, and 2 masses, as described in Section \ref{methods}). Examples of these model transmission spectra that illustrate the effect of each parameter are shown in Figures \ref{transspectra_allplanets}, \ref{transspectra_3planets}, and \ref{transspectra_pressure}. 

Figure \ref{transspectra_allplanets} shows example transmission spectra for all 9 planets considered for each composition (Venus-, Earth-, and Titan-based), assuming an Earth-like surface pressure of 1 bar. Both the mass from the empirical Earth-like composition (darker line) and measured mass (transparent line) are shown. The measured mass has a strong effect on the predicted signal size. In particular, for the TRAPPIST-1 planets b, e, f, g, and h, the measured mass is smaller than the mass if the planets all have Earth-like rock/iron composition (see also Figure \ref{planet_masses}); the features in the transmission spectra models assuming the measured mass are therefore significantly larger than those assuming an Earth-like composition.

Figure \ref{transspectra_3planets} shows transmission spectra for three of the planets: GJ 1132b, TRAPPIST-1c, and LHS 1140b. In each panel, the three different compositions are shown. The locations of features strongly depend on the composition of the atmosphere because of the different gases that are present. 

\begin{figure}[b]
 \vspace{0.2in}
 \includegraphics[width=0.5\textwidth]{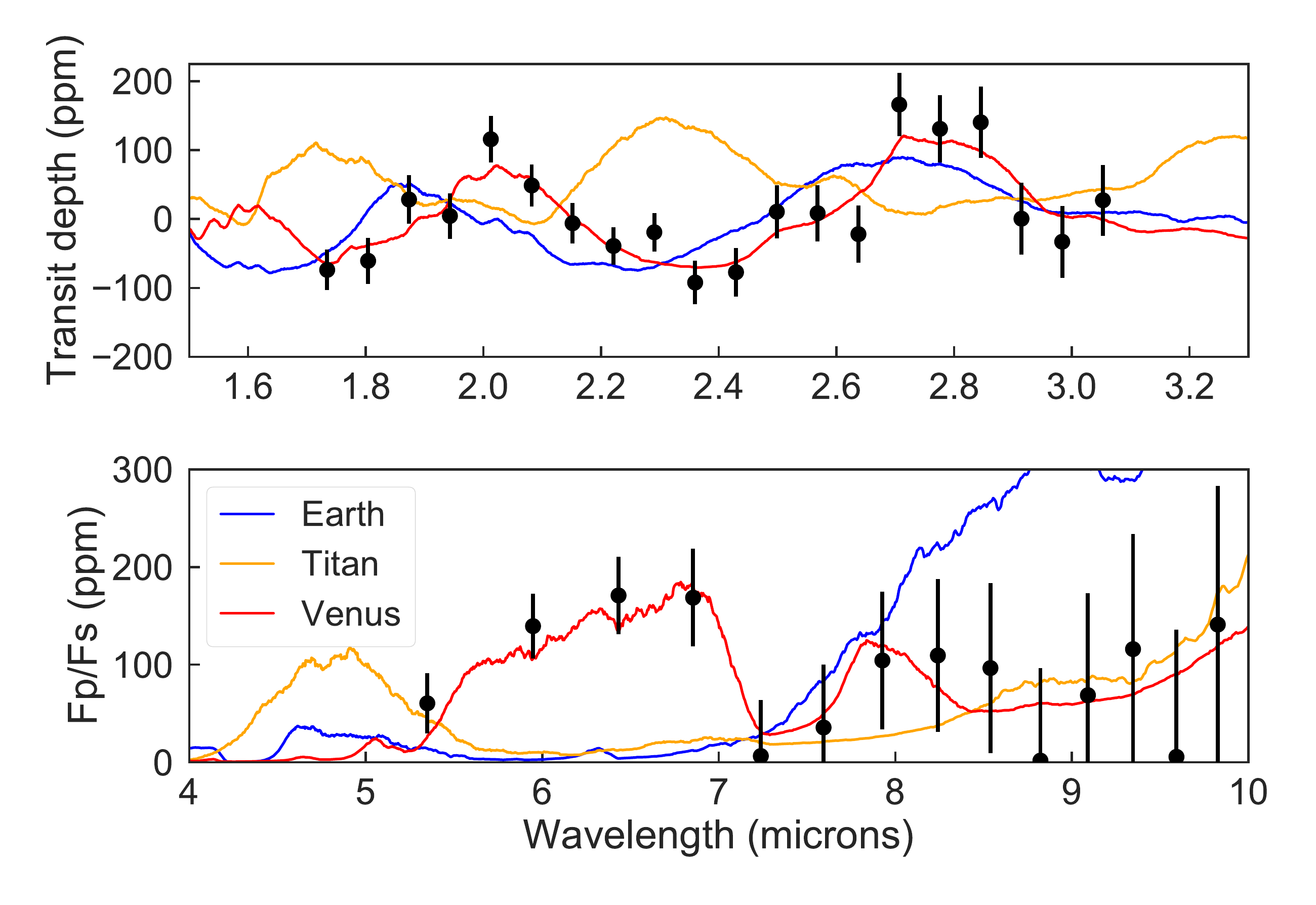}
 \caption{Simulated JWST spectra for TRAPPIST-1b, assuming a 1 bar atmosphere and Venus-like composition (red line). Earth- and Titan-like compositions are shown for comparison (blue and orange lines). The top panel shows the simulated emission spectrum (black points with 1\,$\sigma$ error bars) for 9 eclipse observations with the MIRI/LRS instrument, and the bottom panel shows the transmission spectrum for 6 transit observations with NIRSpec/G235M. These data sets both satisfy the detection criteria outlined in Section\,\ref{jwstresults}.}
\label{jwst-spec}
\end{figure}

Figure \ref{transspectra_pressure} shows the effect of changing surface pressure on the modeled transmission spectrum of TRAPPIST-1b, given a constant Earth-based composition. Transmission spectra of low-pressure atmospheres (<0.01 bar) have small features with flat regions at wavelengths with little gas opacity. This shape indicates that we are probing the surface through the model atmosphere. This result indicates that tenuous atmospheres still have features in transmission spectra. Measuring those features and the flat regions (where we probe the surface) could allow us to constrain surface pressure, assuming that the planet's mass has been more precisely measured, though this may be degenerate with a uniform gray cloud. 

For higher surface pressures, above $\sim$1 bar, the amplitude of features is approximately the same regardless of surface pressure. This result indicates that for higher pressure atmospheres, the transmission spectrum of the planet is extremely insensitive to surface conditions. Detecting at atmosphere in transmission would therefore not probe surface conditions. For example, since surface temperature is a strong function of surface pressure, a transmission spectrum of a planet with a 1 bar atmosphere and temperate surface could be indistinguishable from that of a planet with a 100 bar atmosphere and a very hot surface.

\section{Predictions for \emph{JWST} Observations} \label{jwstresults}

We explored the feasibility of detecting the atmospheres of TRAPPIST-1b-h, LHS 1140b, and GJ 1132b with \emph{JWST}. We used the open-source \texttt{PandExo} tool \citep{Batalha17} to simulate realistic \emph{JWST} spectra that account for e.g. photon, background, and read noise. We made the optimistic assumption that the detectors do not have a noise floor. The actual performance of the detectors of course remains to be seen, but we are heartened by recent analyses of \emph{HST}/WFC3 and \emph{Spitzer}/IRAC data that consistently produces photon noise-limited results down to a precision of 15 parts per million \citep[e.g][]{Kreidberg14, Ingalls16, Line16b}. We note that these predictions do not yet take into account the resolution-linked bias effect identified in \citet{Deming17}, but we expect that this effect would cause a relatively small change in transit and eclipse depth relative to other assumptions (e.g., the large uncertainties in planet mass). 

\subsection{Transmission Spectra}
We considered a range of instrument and disperser combinations to maximize the signal-to-noise for the transit observations. The most intuitive choices were the NIRSpec/Prism and NIRISS/SOSS modes, which both provide high throughput over a wide wavelength range. However, the host stars we are considering are near the saturation limit of these modes, leading to low duty cycles (near 50\%) even if we allow exposure times that reach 100\% of the detector saturation limit. Given that the model transmission spectra have the largest features in the $2-3\,\mu$m wavelength range, we found that the NIRSpec/G235M combination is optimal for detecting spectral features with the fewest transit observations. We therefore simulated data using NIRSpec/G235M with the F170LP filter. We allow a maximum count level of 60\% saturation, which yields a duty cycle of better than 80\% for all three target stars.

We generated mock \emph{JWST} spectra for two subsets of atmosphere models: one assuming zero albedo and the measured planet masses, and the other assuming an albedo equal to 0.3 and masses predicted by the mass/radius relation. For most planets, the first case is optimistic because the TRAPPIST-1 planets are generally less massive than the mass/radius predictions. The lower albedo also results in higher temperature, which increases the amplitude of the features. For each of these cases, we restricted our analysis to a subset of all of the atmospheric pressures (0.1, 1, and 10 bar). As illustrated in Figure \ref{transspectra_pressure}, the shape of the spectra is fairly insensitive to lower or higher pressures.

For each model, we made a sample of 30 simulated JWST spectra, binned them to a resolution of 35, and calculated the significance at which each data set ruled out a flat line. We then determined the number of transit observations required to detect spectral features at 5\,$\sigma$ confidence on average. We do not consider whether we can distinguish different models from each other, as this requires a full atmospheric retrieval analysis; however, with a high signal-to-noise spectrum we can resolve the shapes of different spectral features that will help constrain the dominant absorbing species. The number of transits needed for a 5$\sigma$ detection are listed in Table \ref{observation-summary}. Figure \ref{jwst-spec} shows an example of a simulated transmission spectrum that satisfies our detection criteria for TRAPPIST-1b, assuming a model for the measured planet mass and zero albedo. The simulated data correspond to 6 NIRSpec/G235M transit observations.

The features in the transmission spectra are small (of order 10 parts per million), and thus require repeat observations for a confident detection, ranging from four to over 100 transits for the models we consider. The planet mass is crucial in determining how many observations are needed: the amplitude of spectral features is proportional to the surface gravity, so a factor of two increase in mass corresponds to a factor of four increase in observing time for equivalent signal-to-noise.

For the TRAPPIST-1 system, with the current TTV mass constraints, we could obtain a 5$\sigma$ detection of spectral features for all but one of the planets in fewer than 20 transits (for a Venus composition, 1 bar atmosphere). If instead the masses follow the Earth composition mass-radius relation, all the TRAPPIST-1 planets would require more than 20 transits. GJ 1132b and LHS 1140b have RV masses measured to roughly 30\% precision, and we find that regardless of which mass is assumed (measured or predicted) GJ 1132b can be characterized in 10-20 transits, whereas LHS 1140b requires more than 50 (for surface pressures greater than 0.1 bar).

The atmospheric pressure is also an important factor in determining how many transits are needed. More massive atmospheres are easier to detect than tenuous ones thanks to their larger amplitude features. The number of required transits decreases by a factor of roughly 2 for every factor of 10 increase in pressure (over the range 0.01 to 1 bar). Changing the albedo from 0.0 to 0.3 has less of an effect, with higher albedos increasing the number of transits by <10\% (due to the drop in temperature). Our results are also sensitive to the planet's atmospheric composition. For Earth-like compositions, the cooler planets (with equilibrium temperatures less than 250 K) are less favorable targets because water freezes out, requiring roughly twice as many transits as a Venus-like composition at those temperatures. Titan- and Venus-like atmospheres are less sensitive to changes in temperature because their strongest absorbing species (CH$_4$ and CO$_2$) remain in the gas phase even at low temperatures.

\begin{deluxetable*}{l r r r r r r}
\tablecaption{ Number of transits or eclipses required to detect a Venus-like atmosphere\tablenotemark{a} \label{observation-summary}}
\tablehead{
   \colhead{Planet}  & \colhead{Emission} & \colhead{Emission} & \colhead{Emission} &\colhead{Transmission} &\colhead{Transmission} &\colhead{Transmission}\\  
   \colhead{\,}  & \colhead{P = 0.1 bar}  & \colhead{P = 1.0 bar} & \colhead{P = 10.0 bar} & \colhead{P = 0.01 bar} & \colhead{P = 0.1 bar} & \colhead{P = 1.0 bar}}
\startdata
TRAPPIST-1b	&	6 (11)	&	9 (18)	&	17 (30)	&	23 (89)	&	11 (40)	&	6 (21)	\\
TRAPPIST-1c	&	19 (37)	&	29 (58)	&	48 (92)	&	--	    &	73 (50)	&	36 (25)	\\
TRAPPIST-1d	&	--		&	--	    &	--	    &	59 (--)	&	25 (46)	&	13 (24)	\\
TRAPPIST-1e	&	--		&	--	    &	--	    &	15 (--)	&	6 (66)	&	4 (71)	\\
TRAPPIST-1f	&	--		&	--	    &	--	    &	73 (--)	&	27 (92)	&	17 (54)	\\
TRAPPIST-1g	&	--		&	--	    &	--	    &	36 (--)	&	15 (--)	&	10 (76)	\\
TRAPPIST-1h	&	--		&	--	    &	--	    &	16 (--)	&	6 (90)	&	4 (56)	\\
GJ 1132b	&	2 (2)	&2 (3)	    &	3 (6)	&	27 (38)	&	13 (20)	&	11 (13)	\\	
LHS 1140b	&	--		&	--	    &	--	    &	--	    &	-- (96) &	-- (64) \\
  \enddata
  \tablenotetext{a}{The detection criteria are (1) for transmission spectra, the simulated data must rule out a flat line at 5$\sigma$ confidence on average, and (2) for emission spectra, the band-integrated secondary eclipse must be detected at 25$\sigma$. We base our calculations on models with a Venusian composition, zero albedo, and planet mass equal to the measured values from TTVs or RVs. For the case in parentheses, we use an albedo of 0.3 and planet mass predicted by the theoretical mass/radius relation. The $-$ mark denotes cases where over 100 transits or eclipses are needed.}

\end{deluxetable*} 

\clearpage

\begin{figure}[h]
 \includegraphics[width=3.2in]{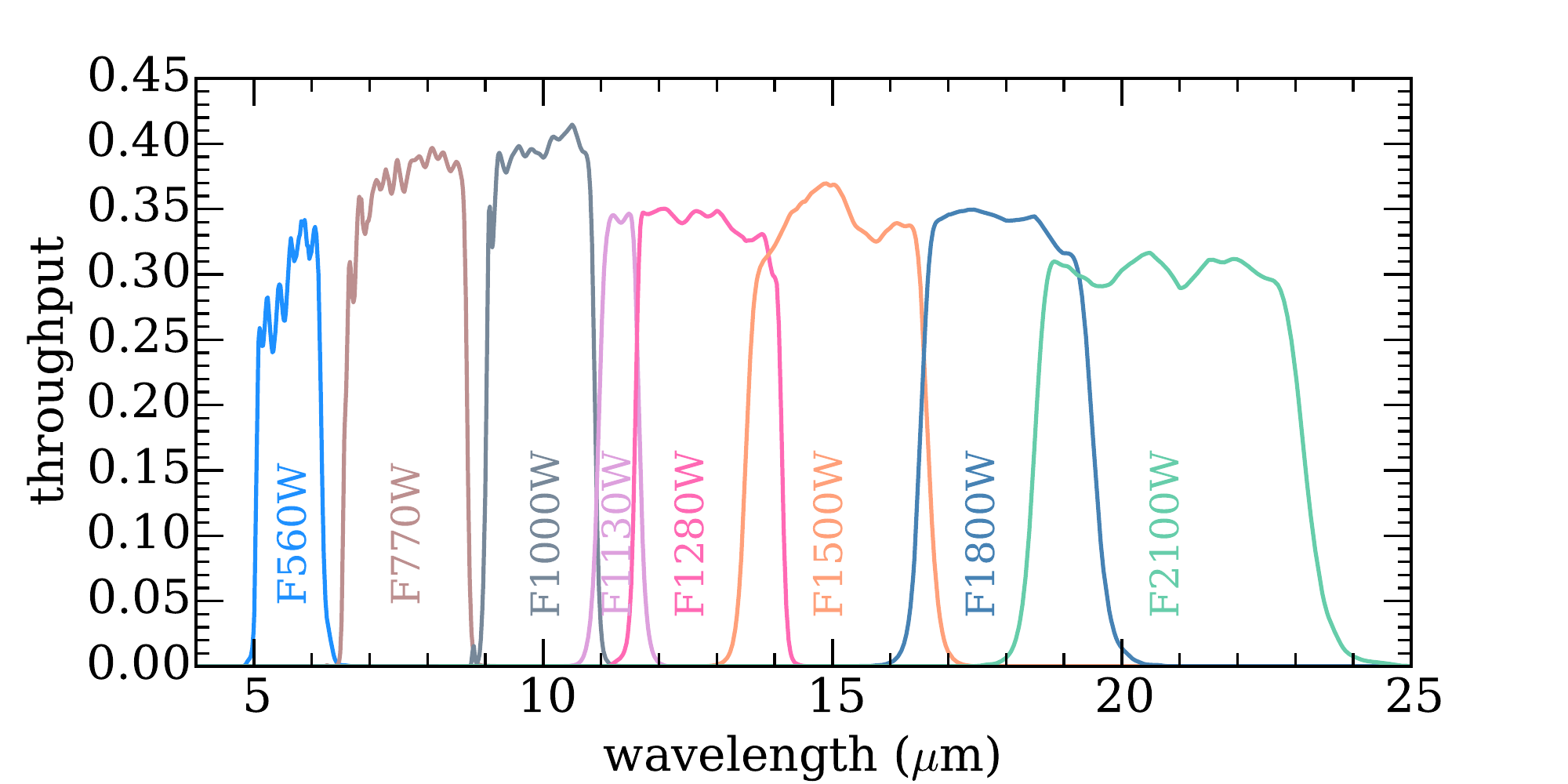}
\includegraphics[width=3.2in]{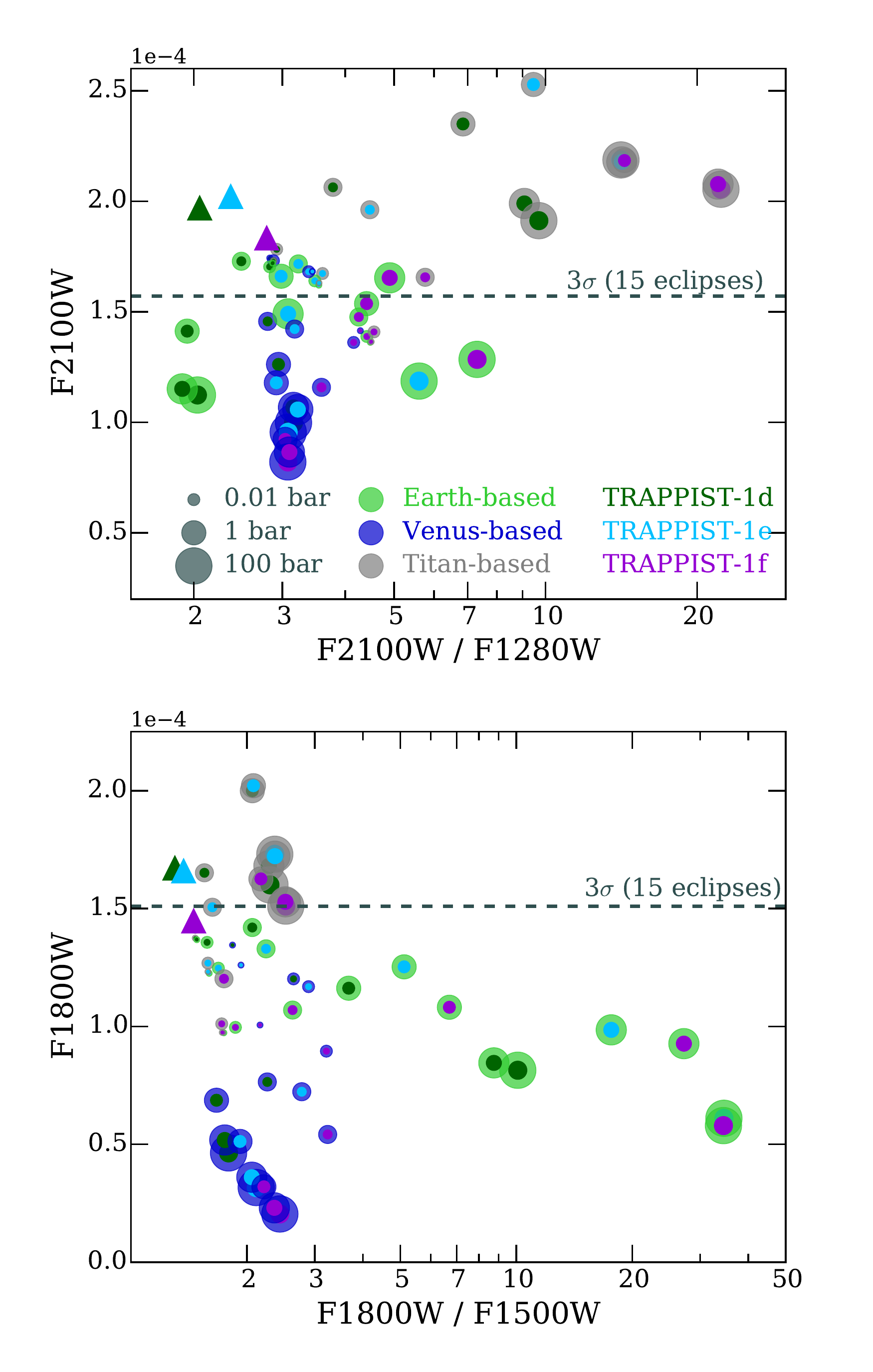}
 \caption{The top panel shows 8 of MIRI's photometric filters and the bottom two panels show eclipse depths in selected MIRI filters for three of the temperate TRAPPIST-1 planets. The x-axis shows the ratio of eclipse depth between two filters and the y-axis shows the depth in a single filter. The top panel shows the F2100W and F1280W filters, while the bottom panel shows F1800W and F1500W. The color of the inner point represents which planet is plotted (TRAPPIST-1d, TRAPPIST-1e, and TRAPPIST-1f); the color of the outer annulus represents the composition of the atmosphere (Earth-, Venus-, or Titan-based composition), and the size of the marker scales with the surface pressure. The three triangular markers in each plot indicate the eclipse depths of an atmosphere-free model assuming zero albedo and no heat distribution, approximated as a blackbody. Thin atmospheres cluster together in this color-magnitude-like space, while the thicker atmospheres cluster according to their compositions. The approximate eclipse depth detectable (3$\sigma$) with 15 eclipses is shown as a dashed line.}
\label{color-mag}
\end{figure}

\subsection{Thermal Emission}

\subsubsection{MIRI/LRS Spectroscopy}
We also simulated thermal emission measurements with the MIRI/LRS instrument. The planets we consider have cool temperatures, so MIRI's mid-infrared wavelength coverage is best for detecting their emitted light. We used \texttt{PandExo} to simulate MIRI spectra, using the LRS slitless spectroscopy mode. We set the maximum full well to 60\% to achieve a duty cycle of better than 90\% for all the observations. We again assumed zero noise floor.

For the simulated eclipses, we determined the detection significance based on the broadband secondary eclipse depth (integrated over the full $5-12\,\mu$m bandpass). We then calculated the number of observations required for a 25$\sigma$ detection of the secondary eclipse. This signal-to-noise ratio is sufficient to resolve spectral features across the bandpass, as illustrated in Figure\,\ref{jwst-spec}. Based on this criterion, we determined the number of eclipses needed for a Venus-like composition and two different albedos: 0 and 0.3. The results are listed in Table\,\ref{observation-summary}. The hottest planets we consider, GJ 1132b and TRAPPIST-1b, are excellent candidates for thermal emission measurements. For either albedo, they will be accessible with 2-3 secondary eclipse observations, fewer than are needed to detect features in their transmission spectra.  The cooler targets are much more challenging: TRAPPIST-1c requires dozens of eclipses, and the other planets all require over 100 observations, which is unlikely to be feasible over the lifetime of \emph{JWST}.

\subsubsection{MIRI Synthetic Photometry}

Predicted eclipse depths in the MIRI imager bandpasses were calculated for each of the available filters (for wavelengths less than 25 \micron). These filters are shown in the top panel of Figure \ref{color-mag}. For temperate planets, spectroscopy with MIRI will likely be unrealistically expensive, but photometry may be more accessible for some targets.

We calculate the synthetic photometry by integrating the planet's photon flux and the stellar photon flux over each of the filters, using photon-to-electron efficiency curves\footnote{http://ircamera.as.arizona.edu/MIRI/pces.htm}, and using the stellar models shown in Figure \ref{stars}. The model photometry will be available with the model spectra online\footnote{www.carolinemorley.com/models, https://doi.org/10.5281/zenodo.1001033}.

In the bottom two panels of Figure \ref{color-mag} we show a transiting planet version of a color-magnitude diagram, using eclipse depths instead of absolute magnitudes. These plots illustrate that in some filters, the models with different compositions separate into different loci. We show the three temperate TRAPPIST-1 planets that have been suggested to be in the habitable zone (TRAPPIST-1d, e, and f). The size of the marker scales with the surface pressure and the colors of the annuli show the composition. 

The approximate eclipse depths of an atmosphere-free planet with zero albedo and no heat redistribution, approximated as blackbody spectra, are also shown (as triangular markers). These eclipse depths are distinct from those of the model atmospheres because the atmospheric models deviate significantly from blackbodies. These ``bare rock'' models are detectable in the F2100W filter with $<$15 eclipses for the TRAPPIST-1 planets suggested to be in the habitable zone (TRAPPIST-1d,e,f). 

Thinner low surface-pressure atmospheres tend to cluster together because they have similar more blackbody-like spectra (they have shallower eclipses than the ``bare rock'' models because all of the atmospheric models assume planet-wide heat redistribution). Thicker high surface pressure atmospheres have more distinct colors.  The three compositions separate because of the different absorbers in each of the model atmospheres. For example, the Titan-based models are comparatively brightest in the F1800W and F2100W filters, because of their relative lack of absorbers at long wavelengths; however, these models are redder than others in their colors because they are comparatively faint in the F1280W filter where ammonia is a strong absorber. The Earth- and Venus-based models have both have significant CO$_2$ present, which has a very strong absorption feature centered at 15 \micron, making them fainter in the F1800W and F2100W bandpasses; the Venus-based models additionally have SO$_2$ features around 20 \micron. Furthermore, the Titan-based models have a somewhat stronger greenhouse effect than Earth-based models (see Figure \ref{psurf_tsurf}), which increase their apparent brightness at long wavelengths.

We simulated MIRI photometry observations with the \textit{JWST} exposure time calculator (https://jwst.etc.stsci.edu/) to calculate the number of eclipses necessary to detect secondary eclipses at 5\,$\sigma$ confidence for a zero albedo model. We used PHOENIX model atmospheres calculated with \texttt{pysynphot2013} \citep{pysynphot2013} for the stellar spectra and normalized them to the target K-band magnitude. We find that the hottest planets are easily accessible with MIRI photometry: TRAPPIST-1b and GJ 1132b can be detected in 2-10 eclipses per filter for most compositions and in most filters redder than 10 $\mu$m. TRAPPIST-1c is also accessible, generally requiring 5-15 eclipses in each of the three reddest filters. The cooler planets are more difficult but may be detectable in certain cases. If we relax the detection significance to $3\,\sigma$, LHS 1140b, TRAPPIST-1d, TRAPPIST-1e, and TRAPPIST-1f are detectable in 10-20 eclipses per filter in the two reddest filters, but only for Titan- and Earth-based compositions. The advantage of observing these planets photometrically is largely due to the wavelengths available; the planet/star flux ratio is higher at longer wavelengths.

\section{Discussion} \label{discuss}

\begin{figure}[h]
\hspace{-0.6cm}
\includegraphics[width=3.85in]{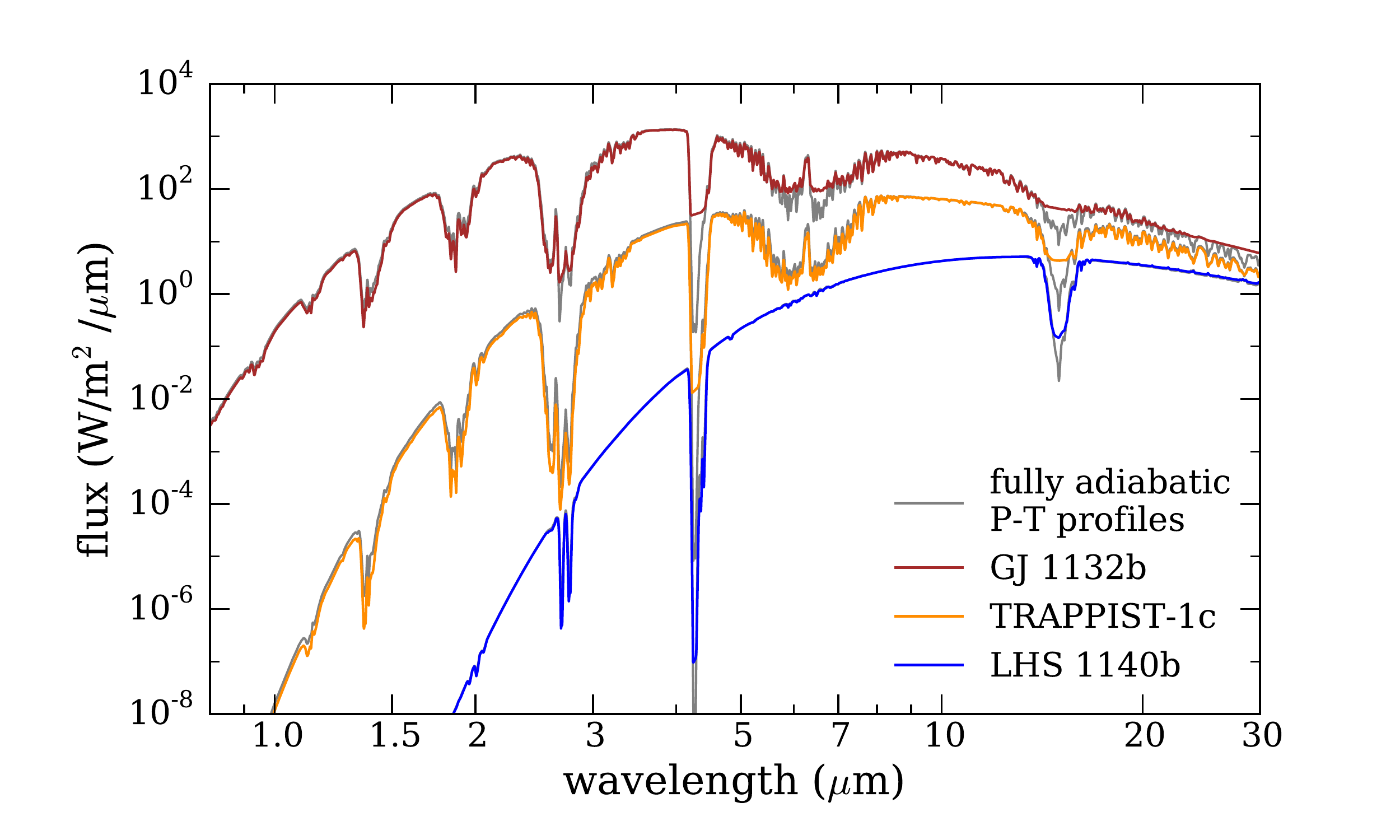}
 \caption{Thermal emission spectra illustrating the effect of the pressure--temperature profile on model spectra. Colors indicate different planets; all models have Earth-based compositions and 1 bar surface pressures. The spectra shown with gray lines have fully adiabatic P--T profiles; the colored lines have P--T profiles with adiabatic deep atmospheres and isothermal profiles above 0.1 bar. Note the deeper absorption features for the fully adiabatic models.  }
\label{adiabats}
\end{figure}

\subsection{Effect of Analytic P--T Profile}

The chosen $P-T$ profile affects the shape of molecular absorption features in the model spectra. For a given composition, the shape of the $P-T$ profile could be calculated using either 1D radiative-convective equilibrium models or 3D climate models. Here, we make simplifying assumptions for the $P-T$ profiles in order to run a wide array of models, as detailed in Section \ref{methods}.


An illustration of how this assumption can change the spectrum is shown in Figure \ref{adiabats}. These models take two limiting assumptions for the $P-T$ profile. Both use dry adiabatic profiles deep in the atmosphere, between 1 and 0.1 bar. In one case, isothermal profiles are assumed above 0.1 bar (colored lines); in the other, a dry adiabat is assumed to extend to the top of the atmosphere (gray lines). The fully adiabatic $P-T$ profile leads to slightly deeper absorption features in regions with strong molecular absorption. The true $P-T$ profile is likely to be between these two assumptions in the absence of an optical absorber that creates a hot (inverted) stratosphere.

\subsection{Effect of Assuming Chemical Equilibrium} \label{disequ}

As described in Section \ref{methods}, we model the molecular abundances in chemical equilibrium. In reality, planetary atmospheres are often not in chemical equilibrium. The actual abundances will be controlled by the source rate and destruction rate for each chemical species. For example, Earth-composition atmospheres in chemical equilibrium have no methane (see Figure \ref{chemistry_synopsis}), but the actual abundance of methane on Earth is about 1.8 ppm (and has increased since pre-industrial times) because natural and anthropogenic sources exceed sinks \citep[e.g.,][]{Dlugokencky11}. In this work we do not seek to run the complex atmospheric chemistry models that would be required to examine this in detail, but we will discuss briefly how photochemistry could affect the compositions and spectra of some of these planets (see  also \citet{Turbet17} for more discussion of photodissociation in the TRAPPIST-1 planets). 

Photodissociation can destroy molecules that would otherwise be present in chemical equilibrium; each molecule has a dissociation energy required for a photon to break one of its bonds. For common molecules, dissociation thresholds are typically in the UV. A molecule must \emph{also} have a high UV cross section to be efficiently photodissociated (i.e. the molecule must have quantum states available at these UV wavelengths). 

\begin{figure}[t]
\hspace{-0.6cm}
\includegraphics[width=3.85in]{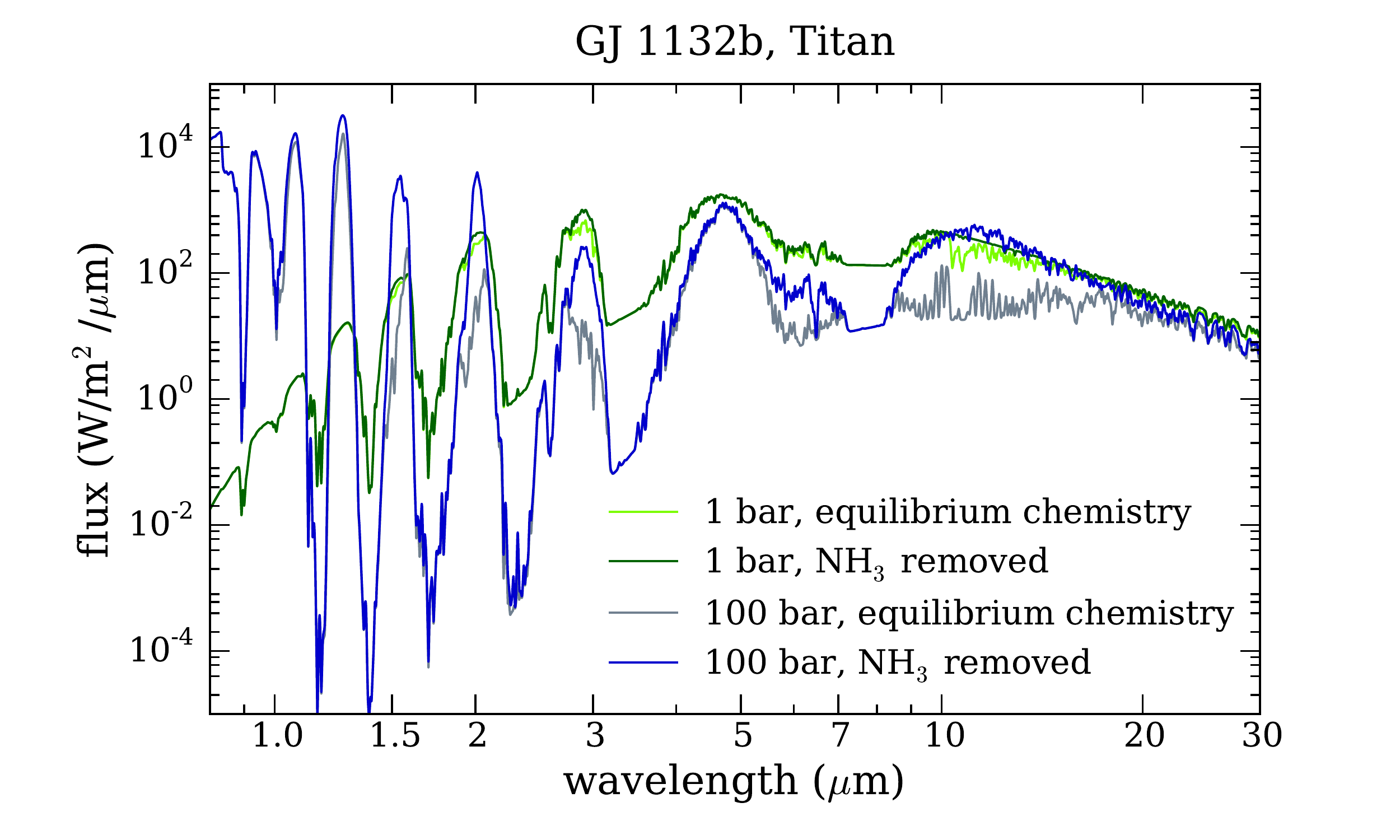}
 \caption{Thermal emission spectra of GJ 1132b illustrating how NH$_3$ absorption affects the spectrum at two different surface pressures. Each set of models have the same P-T profile, and surface pressures of 1 bar and 100 bar respectively. The equilibrium chemistry models assume Titan-based elemental composition; the model with NH$_3$ removed has identical abundances except for NH$_3$. }
\label{removenh3}
\end{figure}

One molecule that is typically readily dissociated in solar system planets is ammonia (NH$_3$). Ammonia has a photodissociation threshold corresponding to photons with $\lambda=0.2637 \mu m$; at these wavelengths, NH$_3$ also has a significant UV cross section \citep{Pierrehumbert10}. After it dissociates, the reactions that form NH$_3$ from the photolysis products are typically inefficient, and so NH$_3$ can be quickly removed from planetary atmospheres. In the models we consider here, NH$_3$ is often in relatively high abundance in the Titan-based atmospheric models and a strong IR absorber (see Figure \ref{chemistry_synopsis} and Figure \ref{opacities}). In Figure \ref{removenh3} we show how removing NH$_3$ would affect the spectra of GJ 1132b. 

The results of photochemistry can increase the abundances of species that would otherwise have very low abundance; for example, in Earth's own atmosphere ozone (O$_3$) is synthesized after oxygen is photodissociated by UV radiation. 

Including the effects of photochemistry will be important for understanding the spectra of these planets, particularly for the planets around active stars with high UV flux. Measuring the UV spectra of the host stars will provide important observationally-driven constraints for these photochemical models.

\subsection{Mass Loss from X-ray/UV Irradiation and Stellar Winds}

Because M dwarfs are are often active compared to solar type stars, especially at young ages, some planets in close orbits around very low mass stars may receive a much higher UV/X-ray incident flux and more intense stellar wind than planets around more Sunlike stars. The survival of primary atmospheres for the lifetimes of these systems has not been empirically determined. Even if the primary accreted atmosphere or first-generation outgassed secondary atmosphere is quickly lost during the M dwarf's youth, planets may outgas secondary atmospheres after the loss of a primordial atmosphere. The balance between mass loss, accretion, and outgassing will control the composition and mass of each planet's atmosphere.

TRAPPIST-1's X-ray luminosity is similar to the current quiet Sun's X-ray luminosity, despite much lower bolometric luminosity, indicating that the planets orbiting it will be subject to much higher X-ray and XUV irradiation than planets of the same temperatures around Sun-like stars \citep{Wheatley17}. This emission is also highly variable; \citet{Vida17} found using K2 data that in addition to quiescent emission, TRAPPIST-1 frequently flares, which may alter the atmospheric chemistry of its planets. \citet{Bolmont16} studied the potential water loss driven by FUV photolysis of water and subsequent XUV-driven escape of hydrogen from the TRAPPIST-1 system, and they found that the inner two planets have likely lost a significant amount of water while the outer planets likely retain more of their water. \citet{Dong17} modeled the stellar wind of the TRAPPIST-1 host star and calculate ion escape rates for the 7 planets, finding that the outer planets of the system are likely capable of retaining their atmospheres. Empirically studying the retention of the atmospheres of the TRAPPIST-1, LHS 1140, and GJ 1132 planets, along with measurements of the high energy emission from the stars themselves, will therefore allow us to constrain the physics of atmospheric mass loss.

\section{Conclusions} \label{conclude}

We have presented a grid of transmission and thermal emission spectra for the 7 planets in the TRAPPIST-1 system and the two Earth-sized planets discovered by the MEarth survey, GJ 1132b and LHS 1140b. These planets span equilibrium temperatures from $\sim$130 K to $\gtrsim$500 K and have measured radii from 0.7 to 1.43 \re. 

In particular, we find that: 

1. Bulk composition of a model planet strongly affects the molecules present and the resulting thermal emission and transmission spectra. We therefore expect that high signal-to-noise observations of terrestrial planets will be sensitive to bulk differences between, e.g., an Earth-like planet and a Venus-like planet. 

2. Thermal emission spectra are intrinsically more sensitive to the surface pressure of a planet than transmission spectra. For surface pressures above $\sim$1 bar, transmission spectra of models with different surface pressures are indistinguishable, whereas the thermal emission spectra of the same model atmospheres are distinct. 

3. Thermal emission spectra are also more sensitive to the equilibrium temperature of a planet than transmission spectra. The temperature of the atmosphere controls at what wavelengths flux emerges in thermal emission, whereas the transmission spectra of models with different temperatures look very similar especially if the planet's mass is not precisely constrained. 

4. Several planets are strong candidates for \emph{JWST} transmission spectroscopy. Assuming the central values of the measured masses are accurate, we could obtain a 5$\sigma$ detection of spectral features with fewer than 20 transits for GJ 1132b and all but one of the TRAPPIST-1 planets for a Venus-composition and a 1 bar atmosphere. However, if the planets have Earth-like bulk compositions, the TRAPPIST-1 planets generally require over 20 transits per planet for a high confidence detection. LHS 1140b is a more challenging target, requiring over 60 transits regardless of which mass is used. Additional observations of the TRAPPIST-1 system are needed to better constrain the planet masses so we can determine the optimal atmosphere characterization strategy.

6. The hottest planets, GJ 1132b and TRAPPIST-1b, are excellent candidates for thermal emission spectroscopy with \emph{JWST}/MIRI. High confidence detection of thermal emission is possible with fewer than 10 eclipse observations. Thermal emission photometry is possible for TRAPPIST-1c in 5--15 eclipses. For the cooler planets (TRAPPIST-1d--f, LHS 1140b) eclipse photometry will be more challenging but potentially possible for certain atmospheric compositions.  

7. The number of transits and eclipses required to detect the atmosphere is sensitive to the atmospheric pressure. Below 1 bar, the number of transits required for a detection increases by roughly a factor of two for each factor of 10 drop in pressure. For emission spectra, planets with lower pressure atmospheres (P$<0.1$ bar) are easier to detect because the blackbody flux from the surface dominates and spectral features are weak; however, these characteristics make it difficult to distinguish between different atmosphere compositions.

As of today, these nine planets represent the most favorable terrestrial worlds for atmospheric characterization. Dedicated campaigns with \emph{JWST} would allow us to detect a variety of different atmospheres and begin to measure the compositions, surface pressures, and surface temperatures of atmospheres around rocky planets outside of the solar system. These will nearly triple the sample size of rocky planets with atmospheres from the $\sim$5 in our own solar system to a broader range of planets around 3 more nearby stars. This increased, more diverse sample of terrestrial planets will allow us to probe the formation and evolution processes of planetary atmospheres, giving us a window into both these new worlds and into our own Earth's atmosphere.

\acknowledgements 
We acknowledge the fruitful and stimulating discussions with Avi Loeb about the TRAPPIST-1 system, and the intellectually vibrant atmosphere of the Institute for Theory and Computation at the Harvard-Smithsonian Center for Astrophysics. We also acknowledge helpful conversations with Robin Wordsworth on modeling terrestrial planets, Michael Line on the calculations of equilibrium chemistry, David Charbonneau on characterizing the planet masses, and Tom Greene on MIRI photometry; their insights all improved the paper. This work benefited from the Exoplanet Summer Program in the Other Worlds Laboratory (OWL) at the University of California, Santa Cruz, a program funded by the Heising-Simons Foundation. This work was performed under contract with the Jet Propulsion Laboratory (JPL) funded by NASA through the Sagan Fellowship Program executed by the NASA Exoplanet Science Institute.


\appendix

To model our transmission spectra, we develop a Python-based radiative transfer code which we introduce here. It utilizes the opacity and optical depth prescription used in the thermal emission code presented in \citet{Morley15} and the three-dimensional geometric path length distribution matrix prescription presented in \citet{Robinson17}. As inputs, the code takes arbitrary pressure-temperature (P-T) and chemical abundance profiles for a planet of specified radius and mass, and it outputs the resultant layer altitudes, thicknesses, optical depths, and the planet's wavelength-dependent transit depth. Gravity and mean molecular weight are calculated in each layer, and altitudes are computed with the assumption of hydrostatic equilibrium.

The optical depth code uses a database of gas species with their corresponding opacity as a function of pressure, temperature, and wavelength. The opacity profiles are based on \citet{Freedman08}, with updates described in \citet{Freedman14}. Collision-induced absorption (CIA) is also included for a number of molecule pairs including N$_2$-N$_2$, O$_2$-O$_2$, CO$_2$-CO$_2$, and N$_2$-CH$_4$ \citep{Gruszka97, Baranov04, Lee16, Bezard90, Wordsworth10, Borysow93, Lafferty96, Hartmann17}, as well as hydrogen and helium CIA opacities \citep{Richard12}. Cross sections are calculated for 1060 P-T grid points that range from $10^{-6}$ bar to 300 bar, and from 75 to 4000 K at 1 cm$^{-1}$ resolution. Wavelengths range from 0.3 to 250 $\mu m$. The summed opacities for each layer are converted to optical depths using the assumption of hydrostatic equilibrium.

We implement the matrix prescription derived in Section 2 of \citet{Robinson17}. In a three-dimensional model atmosphere, incident light rays pass through several layers at different impact parameters, indexed $j$ and $i$, respectively. The path length through any given layer and the number of layers traversed depend on layer thickness $\Delta h_{j}$ and the ray's impact parameter $b_{i}$. As described in more detail in \citet{Robinson17}, a general geometric prescription is used to compute a matrix $P_{i,j}$ which yields the path length through each layer when multiplied by the layer thickness $\Delta h_{j}$. The transmission $t_{\lambda,i}$ through a given impact parameter is represented by the sum

\begin{equation}
    t_{\lambda,i} = EXP\Big(-\sum\limits_{j=1}^{N_{layers}} \Delta \tau_{\lambda,j}P_{i,j}\Big)
\end{equation}
where $\Delta \tau_{\lambda,j}$ is the vertical differential optical depth at a given wavelength and layer. The transit depth over all wavelengths is a sum of the transmission over impact parameters:

\begin{equation}
    \Big(\frac{R_{p,\lambda}}{R_{s}}\Big)^{2} = \frac{1}{R_{s}^{2}}\Big(R_{p}^{2} + 2\sum\limits_{i=1}^{N_{r}}[1-t_{\lambda,i}] b_{i}\Delta b_{i}\Big)
\end{equation}
where $N_{r}$ is the number of impact parameters and $\Delta b_{i}$ is the thickness of each impact parameter gridpoint. 

This new transmission spectrum model is flexible, and it can be used for both giant and terrestrial planets with any number of atmospheric layers and molecular species. It has been validated through comparisons of simple isotherms and Rayleigh scattering atmospheres and yields excellent agreement with a similar model developed by T.D.R.


\end{document}